\documentclass{aa}  
\pdfoutput=1
\usepackage{graphicx}
\usepackage{epstopdf}
\usepackage{txfonts}
\usepackage{ulem} 
\usepackage{soul}
\usepackage{hyperref}
\usepackage{longtable}
\usepackage{lscape}
\newcommand\gaia{\textit{Gaia}\xspace}
\newcommand\gdrtwo{\gaia~DR2\xspace}

\begin{document} 

   \title{Lithium in red giant stars: Constraining non-standard mixing with large surveys in the \gaia era
   \thanks{Based on observations made with the Aur\'elie spectrograph mounted on the 1.52m telescope at Observatoire de Haute-Provence (CNRS), France, with the FEROS spectrograph mounted on the MPI 2.2m telescope at ESO-La Silla Observatory (programmes 072.D-0235A, B; PI: C. Charbonnel), and with the Sandiford Cassegrain Echelle Spectrometer mounted on the Struve telescope at McDonald Observatory, Texas, USA},\thanks{Result tables 1 and 2 are available at the CDS via anonymous ftp to cdsarc.u-strasbg.fr}}

   \author{C. Charbonnel\inst{1,2}, 
   N. Lagarde\inst{3},
   G. Jasniewicz\inst{4},
          P.L. North\inst{5},
          M. Shetrone\inst{6},
          J. Krugler Hollek \inst{6,9},
          V.V. Smith\inst{7}
          \and
          R. Smiljanic \inst{8}
          A.Palacios\inst{4} 
          \and
          G. Ottoni\inst{1}}

   \institute{Department of Astronomy, University of Geneva, Chemin des Maillettes 51, 1290 Versoix, Switzerland
	\email{Corinne.Charbonnel@unige.ch}
	 \and 
	 IRAP, UMR 5277 CNRS and Universit\'e de Toulouse, 14, Av. E.Belin, 31400 Toulouse, France
\and          
Institut UTINAM, CNRS UMR 6213, Univ. Bourgogne Franche-Comt\'e, OSU THETA Franche-Comt\'e-Bourgogne, Observatoire de Besançon, BP 1615, 25010, Besançon Cedex, France 
\and 
UMR 5299 LUPM CNRS Universit\'e Montpellier II, Place Eug\`ene Bataillon, F-34095 Montpellier cedex 05
 \and 
 Institute of Physics, Laboratory of Astrophysics, \'Ecole Polytechnique F\'ed\'erale de Lausanne (EPFL), Observatoire de Sauverny,
           1290 Versoix, Switzerland
\and	
            University of Texas, McDonald Observatory, TX 79734, USA
 \and
            National Optical Astronomy Observatory, Tucson, USA
 \and
            Nicolaus Copernicus Astronomical Center, Polish Academy of Sciences, Bartycka 18, 00-176, Warsaw, Poland
\and
            Twitter, 1355 Market Street, Suite 900, San Francisco, CA 94103, USA
            }
    
\date{Accepted for publication on October 22, 2019}
\authorrunning{C.Charbonnel et al.} \titlerunning{Lithium in red giant stars in \gaia era}
 
  \abstract 
   {Lithium is extensively known to be a good tracer of non-standard mixing processes occurring in stellar interiors.}
      {We present the results of a new large Lithium survey in red giant stars and combine it with surveys from the literature
      to probe the impact of rotation-induced mixing and thermohaline double-diffusive instability along stellar evolution.}
   {We determined the surface Li abundance for a sample of 829 giant stars  with accurate \gaia parallaxes for a large sub-sample (810 stars) complemented with 
      accurate Hipparcos parallaxes (19 stars). The spectra of our sample of northern and southern giant stars were obtained in three ground-based observatories (Observatoire de Haute-Provence, ESO-La Silla, and the Mc Donald Observatory).
We determined the atmospheric parameters ($T_\mathrm{eff}$, $\log(g)$ and [Fe/H]), and the Li abundance. We used \gaia parallaxes and photometry to determine the luminosity of our objects and  
we estimated the mass and evolution status of each sample star with a maximum-likelihood technique using stellar evolution models computed with the STAREVOL code. 
We compared the observed Li behaviour with predictions from stellar models, including rotation and thermohaline mixing. 
The same approach was used for stars from selected Li surveys from the literature.}
   {Rotation-induced mixing accounts nicely for the lithium behaviour in stars warmer than about 4200K, independently
of the mass domain. For stars with masses lower than $2M_{\sun}$ thermohaline mixing leads to further Li depletion below
the $T_\mathrm{eff}$ of the RGB bump (about 4000K), and on the early AGB, as observed. Depending on the definition we adopt, we find between 0.8 and 2.2$\%$ of Li-rich giants in our new sample.} 
   {\gaia puts a new spin on the understanding of mixing processes in stars, and our study confirms the importance of rotation-induced processes and of thermohaline mixing. However asteroseismology is required to definitively pinpoint the actual evolution status of Li-rich giants.} 
  

   \maketitle
%

\section{Introduction}
Along their evolution, stars modify their chemical composition through central and shell nuclear burning coupled to various internal transport mechanisms (atomic diffusion, convection, overshooting, rotation-induced mixing, thermohaline double diffusion, and other kinds of hydrodynamic instabilities). These processes eventually modify the photospheric composition of the stars, in proportions that depend on their initial mass and metallicity, rotation velocity, mass loss, and evolutionary status. 

In the case of low- and intermediate-mass stars (hereafter LIMS), lithium-7 (hereafter Li) is one of the best tracers of the internal mixing mechanisms and of possible associated transport of angular momentum 
\citep[e.g.][]{Deliyannisetal00,TC10}. 
For cool main sequence (hereafter MS) low-mass stars (LMS, i.e., masses below $\sim 1.2$~M$_{\odot}$, which are relatively slow rotators, allowing for the precise determination of their Li abundance), the best constraints are the solar Li photospheric value, which is a factor of $\sim$ 150 lower than the initial meteoritic value \citep[e.g.][]{Lodders_etal09,Asplund_etal09,Caffau_etal10}, the Li abundance of solar-twins \citep[e.g.][]{King_etal97,Martin_etal02,DoNascimento_etal09,Castro_etal11,Monroeetal13,Carlos_etal16}, and the so-called Li dip around effective temperature of 6700~K  that coincides with a sharp variation in the surface rotation rate around mid-F type in open clusters like the Hyades and older \citep[e.g.][]{Wallerstein_etal65,BoesgaardTripicco86,Boesgaard_87,Chen_etal01,Cummings_etal17}. 
Rotation-induced processes coupled to the effects of internal gravity waves are excellent candidates for providing a consistent explanation for the behaviour of Li and that of the surface and internal rotation rate of these MS G- and F-type objects \citep[e.g.][]{TC03,TC05,CCST05}, 
although additional processes like atomic diffusion, mass loss, and magnetic instabilities can play a role 
\citep[e.g.][]{Michaud86,RicherMichaud93,Schramm_etal90,Denissenkov10Sun,Eggenberger_etal12Li}.

Complementary information on transport processes at play in LIMS comes from Li data along the subgiant and red giant branches (hereafter SGB and RGB, respectively). During these evolution phases, surface rotation is usually very low, allowing for a precise exploitation of the spectra. The low Li abundances in SGBs observed well before the occurrence of the first dredge-up \citep[e.g.][]{Balachandran95,Mallick_etal03,Pasquini_etal04,Lebre_etal06,Anthony-Twarog_etal09} and after the end of dilution further up on the RGB \citep[e.g.][]{Brown_etal89} can be explained by rotation-induced mixing that enlarges the Li-free region inside their MS A- or F-type progenitors \citep[e.g.][]{palacios2003}. 
Later on along the RGB, the additional and sudden Li drop observed at the so-called luminosity bump
\citep[e.g.][]{Gratton_etal00,Lind_etal09}  
can be interpreted as resulting from other hydrodynamical instabilities. The most probable one is the so-called thermohaline instability, for which several alternative driving mechanisms have been proposed
(``double diffusive thermohaline mode" induced by $^3$He-burning in the outer wing of the thin hydrogen-burning shell, \citealt{CZ07}, \citealt{ChaLag10}; 
“magneto-thermohaline” mode driven by strong differential
rotation and a poloidal magnetic field, \citealt{Denissenkovetal09}; "cyclonic large-scale vortices" mode where rotation impacts the linear growth of the fingering instability, \citealt{SenguptaGaraud18}; see also phenomenological prescriptions, \citealt{Henkel_etal17}).

Several works in the literature present Li abundance determinations for large samples of field stars. Most of the recent studies focused on dwarf stars, looking for possible connections (which were not found) between Li abundance and the presence of planets (\citealt{Ramirezetal12}, 671 FGK dwarf and subgiant stars; \citealt{DelgadoMenaetal14}, 326 dwarfs; \citealt{DelgadoMenaetal15}, 353 dwarfs). 
Other studies explored the case of post-main sequence stars, starting with the pioneer work of 
\citet[][644 bright G-K giants]{Brown_etal89}, 
followed by \citet[][54 subgiant stars crossing the Hertzsprung gap]{DeLaverny03}, \citet[][154 bright giant stars]{Lebre_etal06}, \citet[][286 G/K giants]{Lucketal07}, \citet[][378 G/K giants]{Liu_etal14}. Recently, large surveys have led to a paradigm shift in terms of sample size. In particular, the AMBRE and GALAH surveys have yielded public catalogues with Li abundances for, respectively, $\sim$~7\,300 and 342\,000 stars covering a large range in metallicity and evolutionary status \citep{GuiglionetalAMBRE16,GALAH2018,DeepakReddy2019}.
A couple of studies have also searched for Li-rich stars from the \gaia-ESO survey \citep{Casey_etal16,Casey_etal19,Smiljanic_etal18} and the LAMOST survey \citep{Cui_etal12,Singh_etal19}. 

In principle, all these data can be used to study the cumulative effects of the various processes that modify the surface Li abundance of LIMS along their evolution. However, robust conclusions require a good knowledge of the  evolution status of individual stars, and a good estimate of their mass. This was the original motivation of the present work, which aimed to derive Li abundances in a volume-limited sample of field giant stars with precise Hipparcos parallaxes (and possibly \gaia measurements), and which are located close to or above the bump on the red giant branch (RGB) and on the early Asymptotic Giant Branch (AGB).
Here we present the results of this observational project, and combine it with other large samples to provide robust and statistically significant constraints on stellar internal mixing processes that are suspected to be in action in LIMS along their evolution.

The paper is organised as follows. We present the selection criteria of our sample stars and determine their luminosity based on Hipparcos or \gdrtwo parallaxes in \S~\ref{sample}. 
The observation runs and the analysis method that led to the determination of the stellar parameters and of the Li abundances are presented in \S~\ref{data_analysis} where we also validate our Li measurements with respect to those derived for common stars in other studies. 
We compare in \S~\ref{section:comp_models_obs} our Li data with the predictions of stellar models including rotation-induced mixing and thermohaline instability computed by some of the co-authors of the present paper \citep{ChaLag10,Lagarde12a}.
We extend the comparison to Li literature data for other samples of giant stars with Hipparcos parallaxes  (\citealt{Lucketal07} and \citealt{Liu_etal14}) or \gaia parallaxes when available \citep[GALAH survey,][]{GALAH2018}.
We present our conclusions in \S~\ref{section:conclusions}.

\section{Sample selection and stellar luminosities with Hipparcos and/or \gaia parallaxes}
\label{sample}

\subsection{Initial sample selection} 
\label{sampleselection}

Our initial goal was to use Li abundances in evolved stars to constrain the cumulative effects of internal mixing processes acting both on the MS and the advanced phases of the evolution of LIMS, in addition to searching for Li-rich giants.  We originally selected a volume-limited sample of stars potentially located on the red giant branch (RGB) and the early Asymptotic Giant Branch (AGB), using their Hipparcos parallaxes, V magnitude, and B-V and V-I colours. 
This sample was extracted from the first Hipparcos catalogue \citep{HIP1}, requiring that the parallax be larger than 5mas, with a relative error of less than 10~$\%$.
Bolometric corrections were derived from V-I given in the same catalogue to estimate the luminosity of the candidates. All the stars in the sample are brighter than V=8. The RGB stars were selected using V-I$>$0.9 and 1.2$<$log(L/L$_{\odot}$)$<$2.3, and the possible AGB stars using B-V$<$1.5 and log(L/L$_{\odot}$)$>$2.6. 
We favoured these two regions of the Hertzsprung-Russell diagram (HRD) because they correspond to the location of the RGB bump and that of the early-AGB and they were thought to host most of the so-called Li-rich giants \citep[][see \S~\ref{subsection:Li-richstars}]{CharbonnelBalachandran2000}. 
To derive statistically significant features, we observed all the presumed early-AGB stars and 2/3 of the presumed RGB stars selected that way. Hence the total number of stars for which we carried out spectroscopic observations is 829  
(731 probable RGBs and 98 probable AGBs). 

\begin{figure} 
	\centering
        \includegraphics[angle=0,width=9cm]{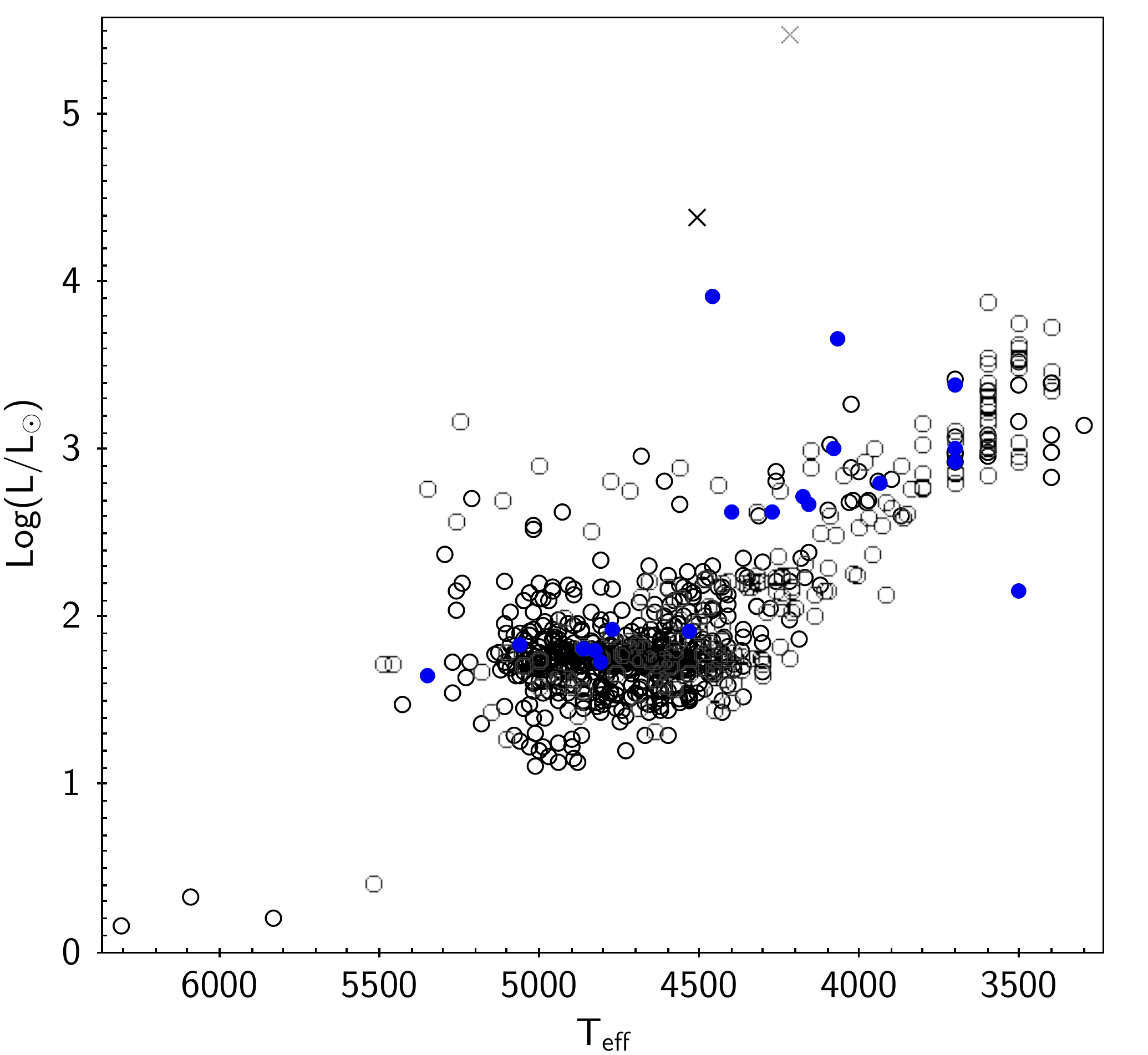}
	   \caption{Positions in HRD of our entire sample.  Stars with \gdrtwo  parallaxes are shown by black open symbols (circles and crosses for parallaxes better or worse than than 10~$\%$, respectively); blue symbols are stars with New Hipparcos parallaxes only (all better than 10~$\%$). Input parameters from the different catalogues used for the determination of the stellar luminosities and effective temperatures are described in \S~\ref{parallaxe} and \ref{subsection:Teff_gravity} respectively}
	\label{HR_all}
\end{figure}

\begin{figure}
    \centering
    \includegraphics[angle=0,width=9cm,trim=0cm 0cm 2cm 2.2cm, clip=true]{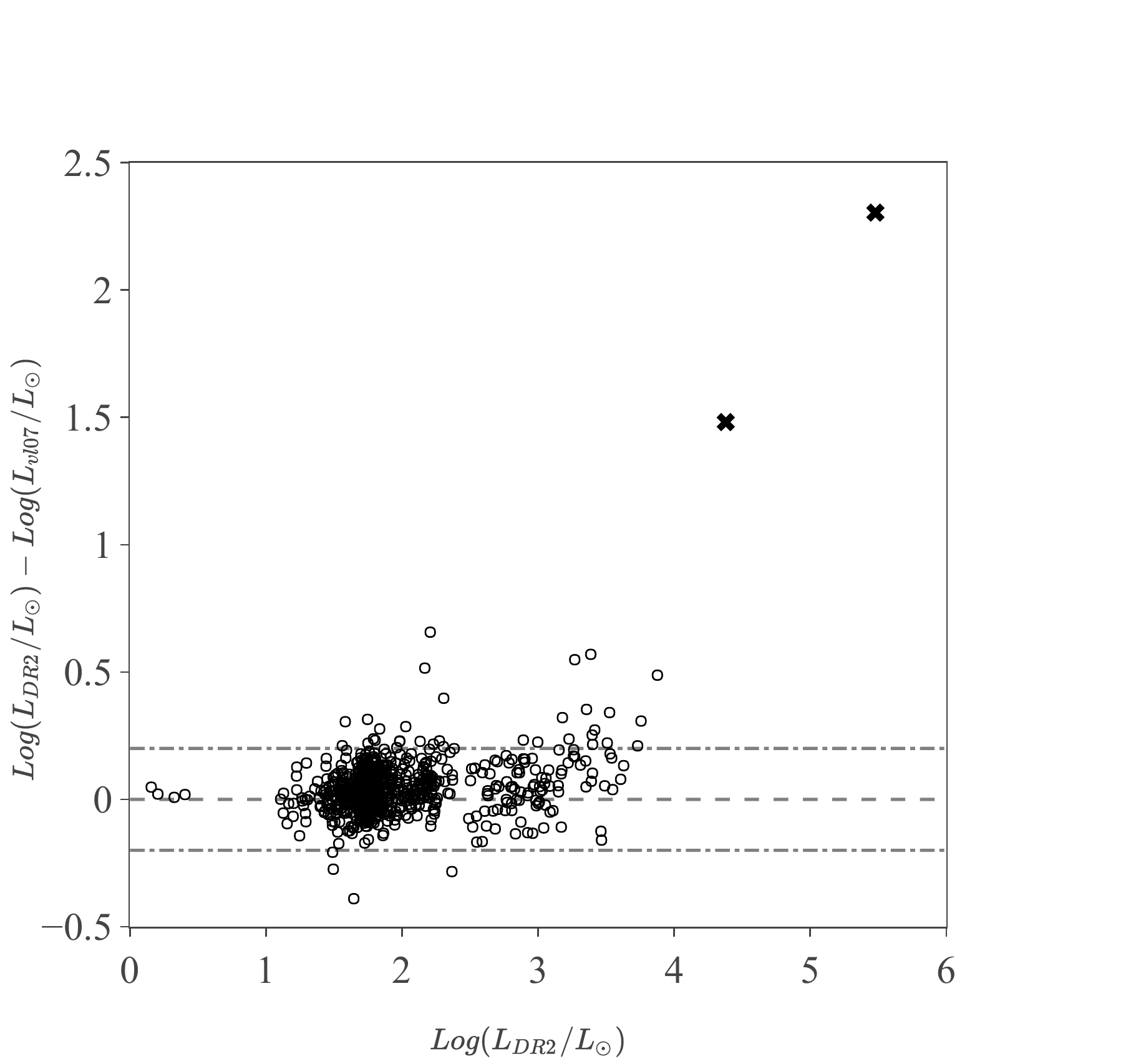}
    \caption{
    Comparison between two sets of luminosity values derived for our sample stars using \gdrtwo or New Hipparcos parallaxes as described in \S~\ref{parallaxe}. Circles and crosses correspond to stars with  \gdrtwo parallaxes better or worse than than 10~$\%$, respectively.
    The dashed-dotted lines represent a difference of 0.2~dex}
    \label{fig:Lcompa}
\end{figure}

\subsection{Determination of stellar luminosities and final sample based on \gdrtwo}
\label{parallaxe}

We computed the luminosity of our sample stars using parallaxes from \gdrtwo  \citep{BrownetalGaiaDR2_2018} when available (810 stars out of 829), or from the New Hipparcos catalogue \citep{HIP2} for the remainder (19 stars). We adopted V magnitude from \citet{Evans2018} when available (800 stars), from Tycho-2 (25 stars) or \citet[][4 stars]{HIP2} catalogues otherwise. In all cases we applied the bolometric correction relation of \citet{Flower1996}. 
We provide the corresponding luminosity value for each individual star in Table~\ref{Observations_table}, together with flags indicating the origin of the parallax and of the V magnitude (flag$_L$). 
We adopted these values of the luminosity for our analysis and use them to plot the 
positions of our sample stars in the HRD 
(Fig.~\ref{HR_all}, \ref{HR_LI_all}, \ref{fig:massesoursample}, \ref{fig:HR_Li_tracks_Zsun} and
\ref{fig:Lirich}). 
In Fig.~\ref{HR_all} we differentiate the stars that have \gdrtwo parallaxes better than 10~$\%$ (805 out of 810; 5 stars have errors between 10 and 20~$\%$, and 2 outliers have errors larger than 70~$\%$); the 19 stars with no \gdrtwo parallaxes have New Hipparcos parallaxes better than 10~$\%$. 
The apparent lack of stars with Log(L/L$_{\odot})$ between $\sim$ 2.3 and 2.6 results from our sample selection.

For a consistency check, we also computed the luminosity using the parallaxes and V magnitudes from \citet{HIP2} for the entire sample, assuming a bolometric correction relation of \citet{Flower1996}. The corresponding luminosity values are also given in Table~\ref{Observations_table} and compared to the previous ones in Fig.~\ref{fig:Lcompa}. Except in a few cases, the agreement between the two sets of luminosity is reasonable  (e.g. within $\pm 0.2$ dex). We refer to the original \gdrtwo papers \citep{Babusiaux_etal18_GaiaDR2HRD,BrownetalGaiaDR2_2018,Lindegren_etalGaiaDR2_2018} for general discussions on the differences between their catalogue and that of  \citet{HIP2}.

\section{Observations and spectra processing}
\label{data_analysis}

\subsection{Observations and data reduction}
\label{observations}

Several observing runs were performed to secure the original data set built for the present study with the selection criteria described in \S~\ref{sample}.

Three observing runs were performed at the
Observatoire de Haute Provence, from 8 September to 15 September 2003 (visitor mode), from 17 January to 19 January, 2004 (service mode), and 
from 16 January  to 16 February, 2004, service mode).
High-resolution spectra were obtained with the Aur\'elie spectrograph
\citep{GBK94} attached to the 1.52 m telescope. Since Aur\'elie has no
cross-disperser, we selected the spectral range centred on the Li $\lambda670.8$\,nm line and about 20\,nm wide. 
The Aur\'elie spectrograph uses a cooled 2048-photodiode detector forming
a 13 $\mu$m pixel linear array. The entrance spherical diaphragm
is 3$\arcsec$ wide. For the spectral range centred on $\lambda670.8$\,nm,
the grating \#5 with 1200 grooves mm$^{-1}$ was used, giving
a mean dispersion of $0.26$\,nm\,mm$^{-1}$ at order 2, a resolution $R=70\,000$
and a spectral coverage of about $7$\,nm. The order-separating filter was
the OG515. 
At the beginning, end, and middle of each night, thorium lamps were
observed for wavelength calibration, while at the beginning and end of each
night, a series of nine flat-fields were obtained using an internal lamp
(Tungsten), as well as a series of five offsets. These spectra were
reduced using standard MIDAS procedures.
The exposure times were adjusted to obtain a signal-to-noise ratio (S/N) higher than $\sim 150$ per pixel of the extracted spectrum in the Li line region.

Two observing runs were performed in visitor mode at ESO-La Silla, Chile, from 3 January to 6 January, 2004, and from 5 March to 8 March, 2004.
High resolution ($R\simeq 43\,000$) echelle spectra were obtained with the
FEROS spectrograph attached to the MPG/ESO 2.2m telescope. Each spectrum covers
the whole visible range and even beyond, between about $400$ and $900$\,nm. 
All spectra were reduced using the ESO pipeline FEROS-DRS which runs within MIDAS.
The typical S/N obtained is $150-200$ per pixel of the extracted spectrum in the Li region.

A total of 92 stars were observed at the McDonald Observatory with the Otto
Struve Telescope using the Sandiford Cassegrain Echelle Spectrometer
(SES) from August 2003 to May 2004 over the course of seven observing
runs.  The \#9 slit (1.1" x 3.5") was used to yield a two pixel
resolution of $60\,000$.  After each star's observation, a ThAr exposure and a flat field
were taken for wavelength calibration purposes. The wavelength region
spans from 618 to 815\,nm in the red portion of the spectra.  The data
were reduced using the standard echelle routines in IRAF.


\subsection{Determination of effective temperature and gravity}
\label{subsection:Teff_gravity}

\begin{figure}
    \centering
    \includegraphics[angle=0,width=9cm,trim=0cm 0cm 2cm 2.2cm, clip=true]{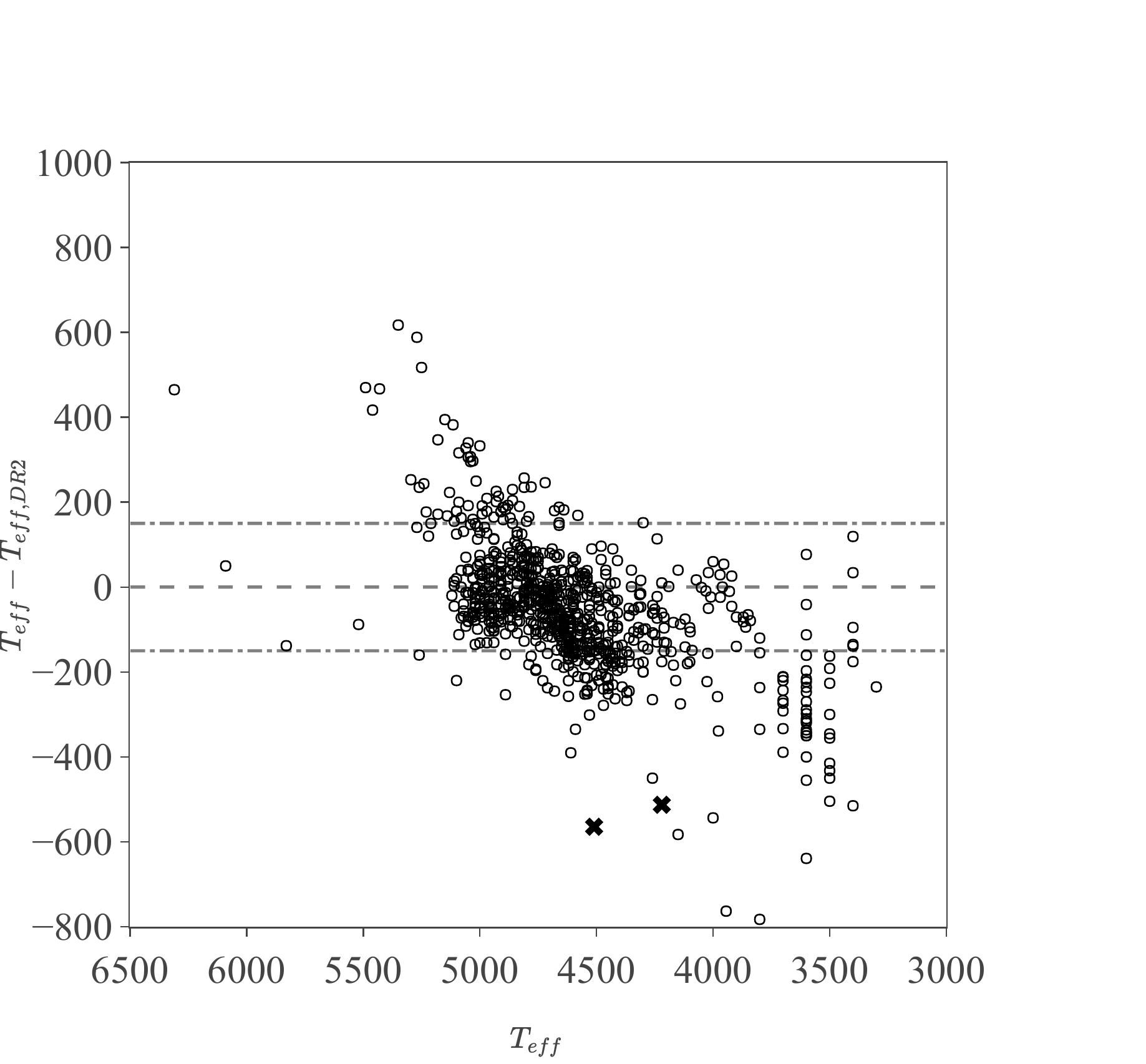}
    \caption{Comparison between effective temperature determined here and that published in \gdrtwo for our sample stars (same symbols as in Fig.~\ref{fig:Lcompa}).  
    Dashed-dotted lines represent a difference of 150~K in T$_{\rm{eff}}$}
    \label{fig:Teffcompa}
\end{figure}

\begin{figure} 
	\centering
       \includegraphics[angle=0,width=9cm]{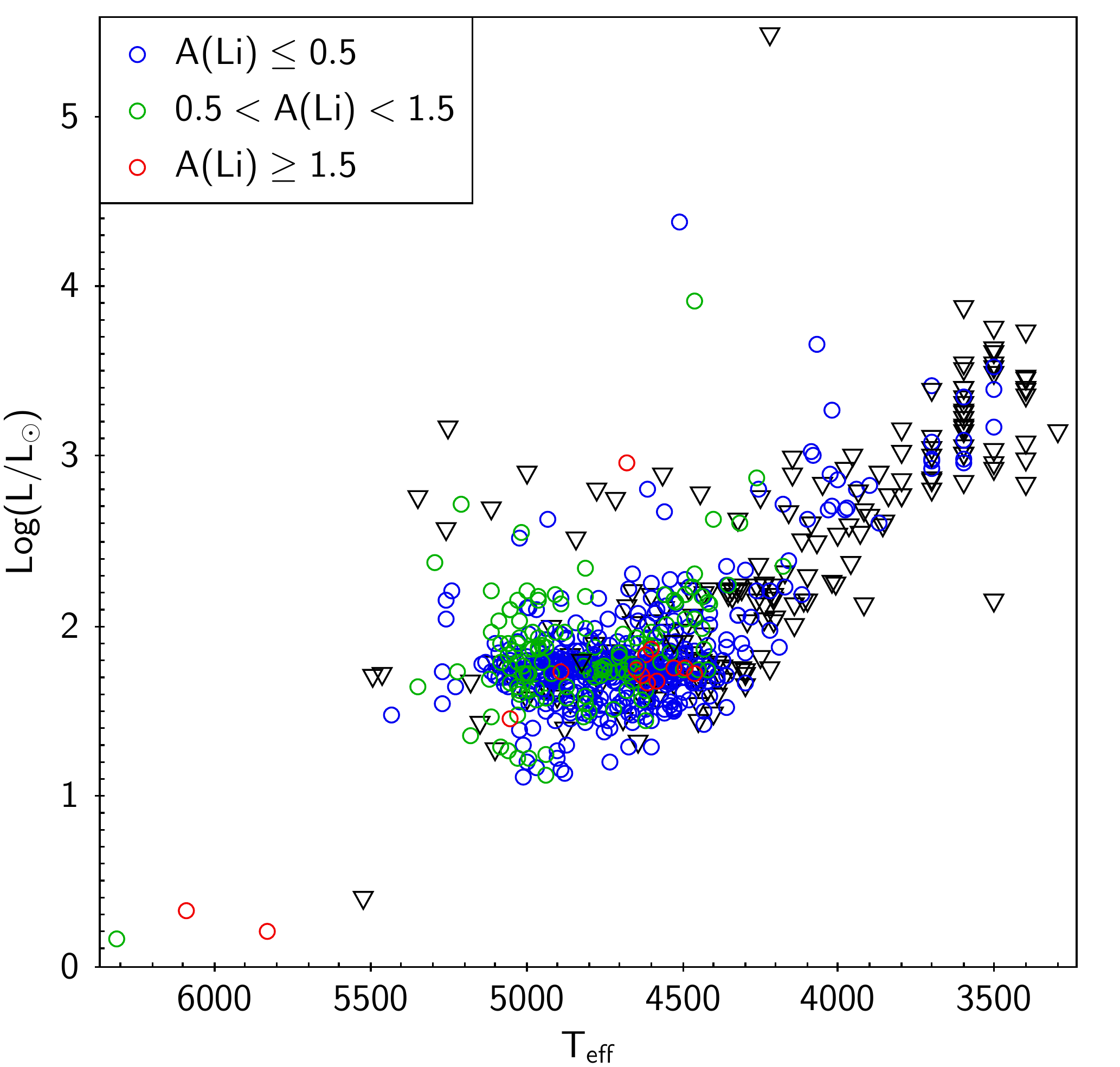}
	  \caption{Indication of  range in Li content (LTE; colour-labelled; circles are for abundance determinations and triangles are for upper limits) derived in this work for all individual stars of our sample with their positions in HRD. 
      Luminosities are computed as in Fig.~\ref{HR_all} (see text for more)} 
	\label{HR_LI_all}
\end{figure}

All spectra coming from the OHP, ESO, and MacDonald observatories were processed in the same way.
Since spectra analysis was done well before \gdrtwo, we took the required quantities listed below from the Hipparcos catalogue.

For each star, the effective temperature $T_\mathrm{eff}$ was determined
by using both B-V and V-I colours 
from the New Hipparcos catalogue 
\citep{HIP2} 
and the  colour-temperature relations for F-K giants from
\citet{Houda}. For the coolest stars ($T_\mathrm{eff} \leq4000K$) we used \citet{Alonso99}.
Reddening was low for typical distances of the order of 100\,pc and can be neglected as
done by \citealt{Lucketal07}. 
The error on $T_\mathrm{eff}$ was estimated from an interpolation processing
in the Houdashelt tabular data assuming $2.0\leq\log g\leq3.0$.
The stellar effective temperatures and the corresponding errors are given in Table~\ref{Observations_table} and they are used across different figures throughout the paper. We compare our values for the effective temperature with those of \gdrtwo  \citep{BrownetalGaiaDR2_2018} in Fig.~\ref{fig:Teffcompa}. As expected, the largest differences are found among the coolest stars with effective temperature below $\sim$ 4000~K (see e.g. \citealt{Andrae_etal18} and \citealt{Arenou_etal18}).  
The cut at $4000K$ mentioned above in our $T_\mathrm{eff}$ calibration procedure also reinforces the observed discrepancies in Fig.~\ref{fig:Teffcompa}. 

For the spectra analysis we consider bolometric magnitude $M_\mathrm{bol}$ and gravity $\log(g)$ 
as given by 
$M_\mathrm{bol} = V - 3.1\,E(B-V) + BC(V) + 5\log(\pi) + 5$
and 
$\log(g) = \log(M) +0.4M_\mathrm{bol} + 4\log(T_\mathrm{eff})
-12.505$
using the 1997 Hipparcos parallax $\pi$ and 
$M=2M_{\sun}$ for 
all G-K giant stars; 
the error on $\log(g)$ comes essentially from the error on mass.

\subsection{Determination of [Fe/H] and of lithium abundances}
\label{subsection:abunddetermination}

All spectra and calibration files were processed with
the MIDAS package.
For each star, a series of synthetic spectra with effective
temperature $\log(T_\mathrm{eff})$ and gravity $\log(g)$  
derived as described in \S~\ref{subsection:Teff_gravity} 
and with various metallicities ($-0.6\lid [Fe/H]\lid0.2$) were calculated
between 665 and 675nm  
with tools from the Uppsala Stellar Atmospheres Group \citep[][]{Gustafsson_etal2008}. 
The MARCS model atmospheres used here 
are line-blanketed local thermodynamic equilibrium 
(LTE) models that include spherical symmetry, the opacity sampling treatment of line
opacities, a chemical equilibrium extended to a few hundred molecular
species, opacity for TiO and $\mathrm{H}_{2}0$, and 
updates of the continuous and atomic opacities \citep[][]{Gustafsson_etal2008}.
We adopted a uniform micro-turbulent velocity of $2.0$~km\,s$^{-1}$ for all
synthetic spectra.
The model spectra use the line list around 671nm
extensively described by \citet{lines}.
 With the help of the computed
synthetic spectra, some short spectral windows where the
continuum approaches unity were identified in
our observed spectra and used to draw the continuum.
After the normalisation of observed spectra by their continuum,
corrections for radial velocities were performed by using a
MIDAS cross-correlation procedure between observed and synthetic spectra. 
Mean metallicity and lithium abundance are determined
for each star by the best fit between the normalised spectrum 
and synthetic spectra:
[Fe/H] was at first derived from metallic lines  (mostly from about thirty Fe lines between 665 and 675nm), and then 
A(Li)\footnote{A(Li)=log($\frac{X(Li)}{X(H)}\cdot \frac{A_{H}}{A_{Li}}$)+12, with X(Li) the Lithium mass fraction} is derived from the resonance LiI $\lambda670.8$nm line.
The error on the mean metallicity estimated from
the fit of the iron lines found in the vicinity of the Lithium line
is around 0.1~dex. The final adopted uncertainty of the Li abundance
is around 0.2~dex. 

\begin{figure} 
	\centering
        \includegraphics[angle=0,width=9cm]{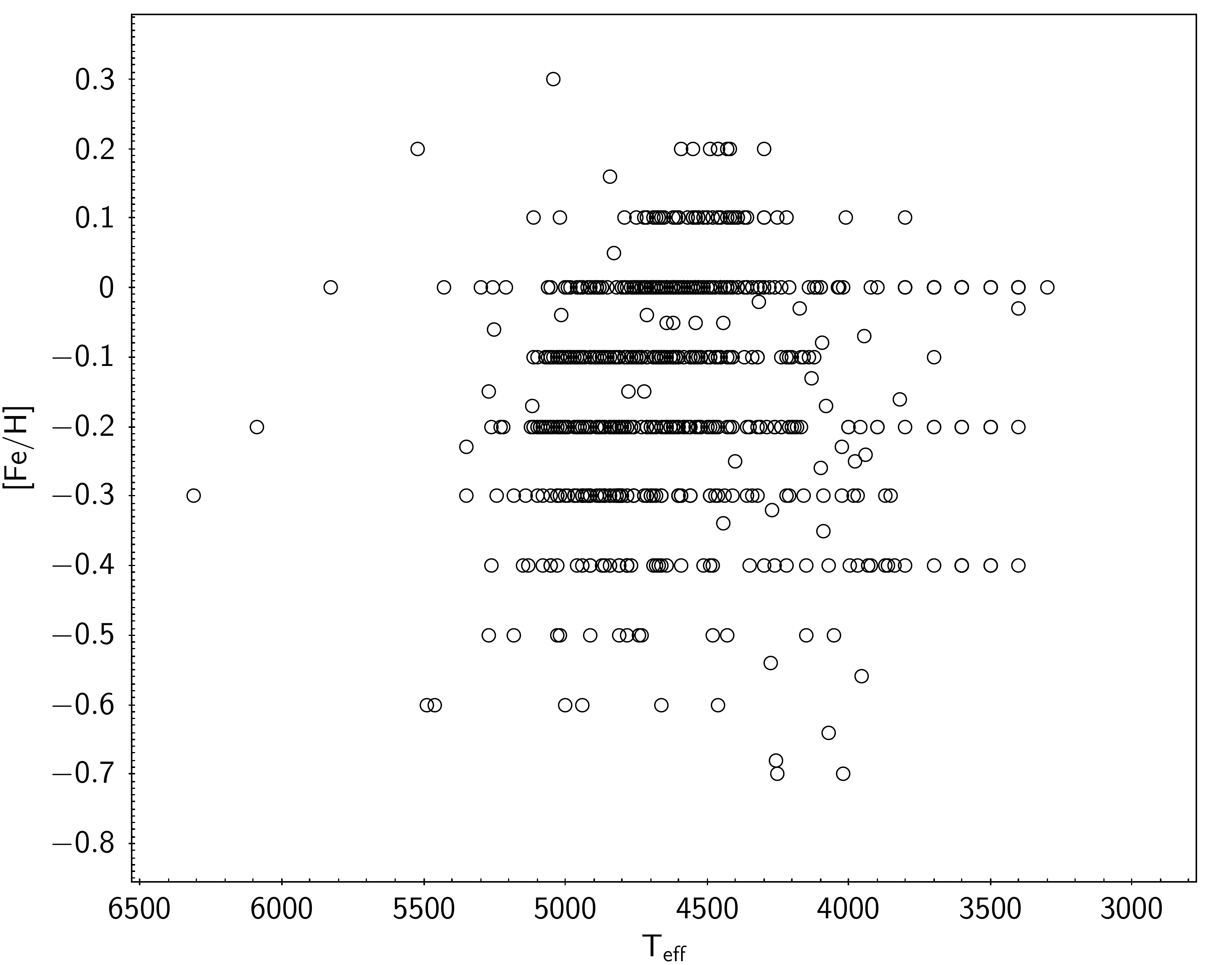}
   \caption{[Fe/H] as function of effective temperature for our sample stars}     
\label{fig:FesurH}
\end{figure}

Abundances of Li and [Fe/H] values are given in Table~\ref{Observations_table}. An indication of the Li range for all individual stars is presented in the HRD in Fig.~\ref{HR_LI_all}.  [Fe/H] values are shown as a function of effective temperature in Fig.~\ref{fig:FesurH}. Respectively 79.2, 20.6, and 0.2~\% of our sample stars have $-0.2 \leq$ [Fe/H] $\leq +0.2$,
[Fe/H]$<-0.2$, and [Fe/H]$> +0.2$ (29~\% of the total have [Fe/H]$=0$). 

The Li abundances given here are computed in LTE. We are aware that non-LTE (NLTE) effects
are not negligible, especially on the \ion{Li}{i}$\lambda 6708$ line used here for
abundance determination. \citet{LAB09} have computed NLTE corrections for giants with
$T_\mathrm{eff}$ between $4000$\,K and $8000$\,K and various metallicities, surface
gravities, and Li abundances. We did not apply these corrections to our LTE abundances,
particularly as we would have had to extrapolate them for stars with $T_\mathrm{eff}< 4000$\,K, that is, for about $10$\% of our sample. Furthermore, the grid
covers Li abundances down to A(Li)$=-0.3$, while our observed abundances or upper 
limits are lower than this value for several tens of stars.
As shown by \citet{Liu_etal14}, the NLTE correction $\Delta_{NLTE}\equiv
A_\mathrm{NLTE}-A_\mathrm{LTE}$ depends above
all on $T_\mathrm{eff}$: while $\Delta_{NLTE}$
amounts to only about $0.1$~dex for $T_\mathrm{eff}\geq 5200$\,K, it reaches about
$0.3$~dex at $T_\mathrm{eff}=4500$\,K. For cooler stars, which represent roughly $1/3$
of our sample, Table~2 of \citet{LAB09} displays NLTE corrections as large as 
$0.33$~dex
for solar metallicity stars with $T_\mathrm{eff}=4000$\,K, $\log g=2.0$, [Fe/H]$=0.0$
and A(Li)$=0.0$ (for a $2$~km\,s$^{-1}$ microturbulence). For higher effective temperatures,
the corrections are smaller but always positive as long as A(Li)$\leq 2.7$. In other words,
our LTE Li abundances underestimate the true abundances by a few tenths of a dex at most.
We will comment this briefly in \S~\ref{section:comp_models_obs}.

\subsection{Comparisons with other studies}
\label{section:comparison}

\begin{figure} 
	\centering
        \includegraphics[angle=0,width=7cm]{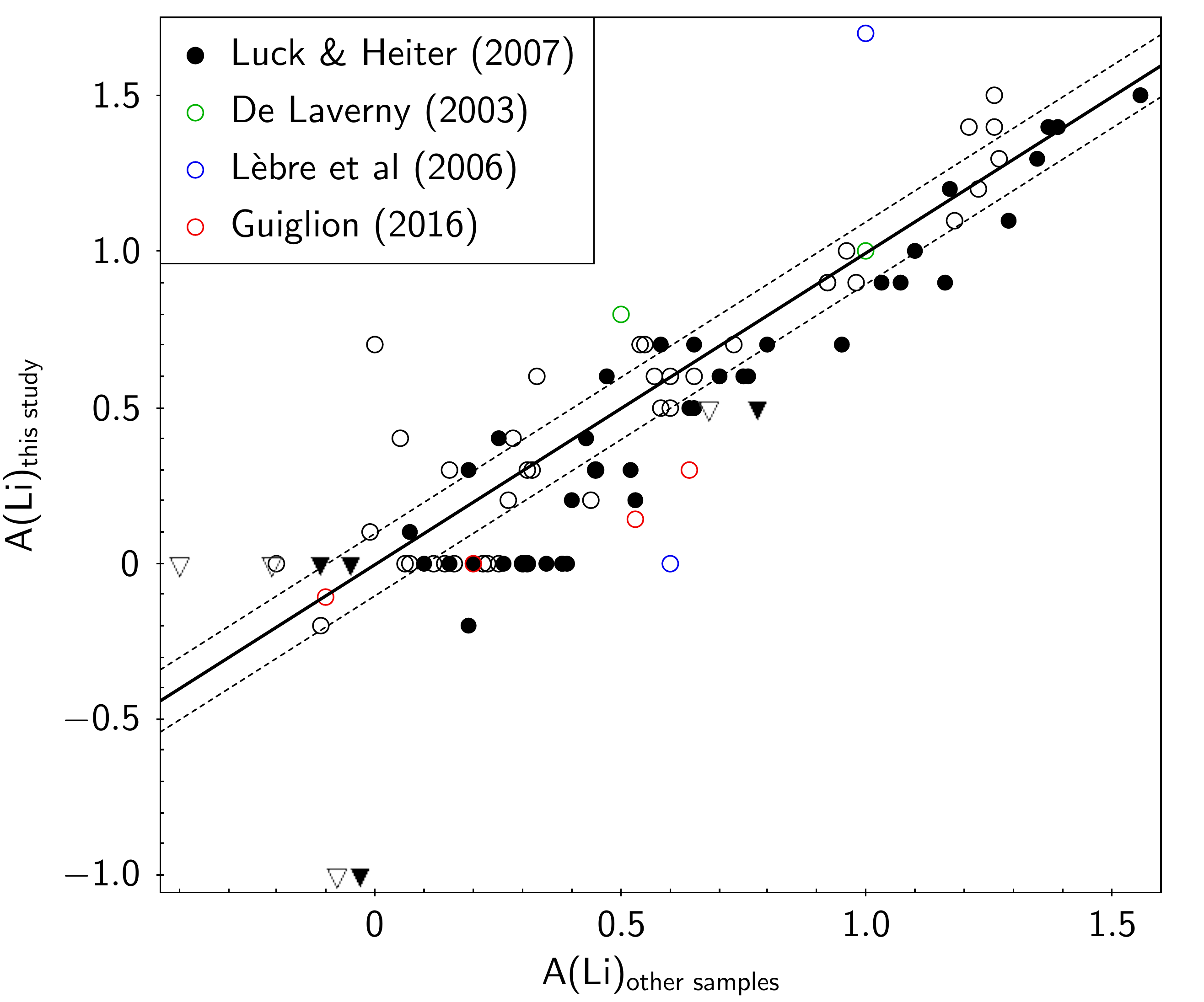} 
        \includegraphics[angle=0,width=7cm]{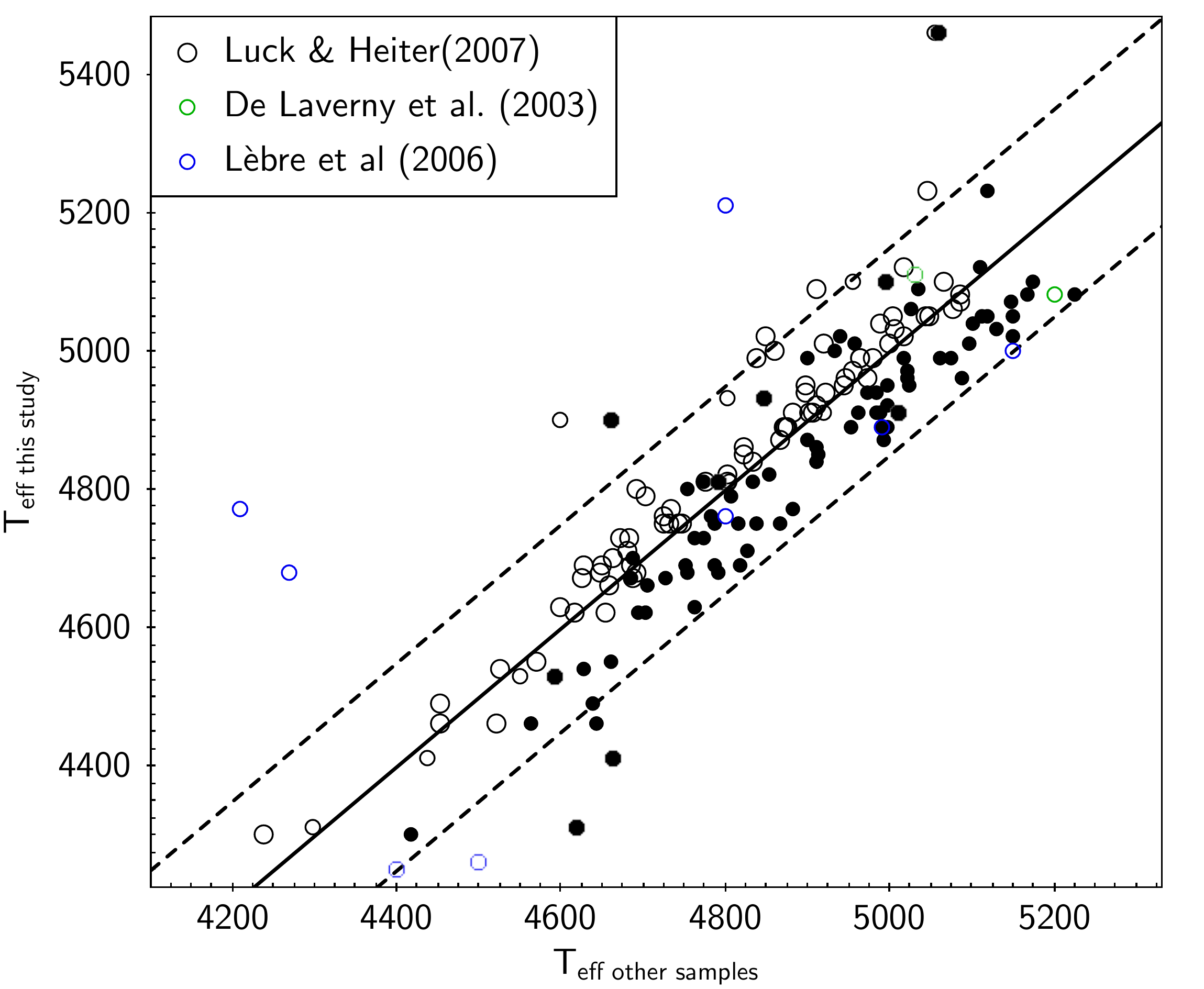} 
        \includegraphics[angle=0,width=7cm]{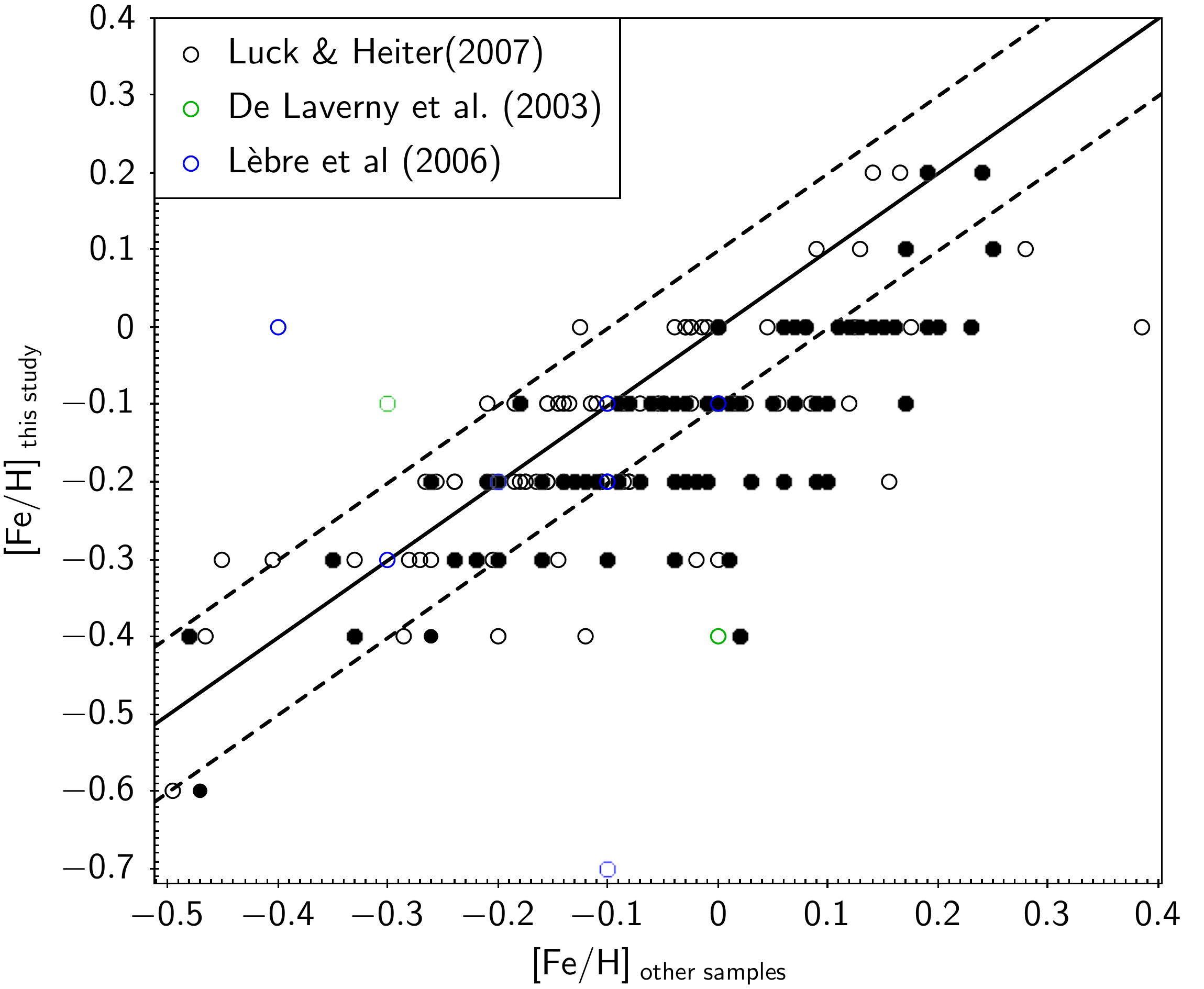}              
	  \caption{
	  (Top) Comparison between LTE Li abundances determined in this work and LTE Li abundances published in studies of \citet[][green]{DeLaverny03}, \citet[][blue]{Lebre_etal06}, \citet[][black]{Lucketal07}, and \citet[][red]{GuiglionetalAMBRE16}. 
	  We exclude the stars with upper limits for Li published in the literature, but we include our upper limits in the case of real determinations by the other studies. (Middle and bottom) Comparison with effective temperature and [Fe/H] values of same studies when available. For the sample of \citet[][]{Lucketal07}, we show their photometric and spectroscopic effective temperatures and corresponding Li and [Fe/H] values (open and full black circles). Dotted lines represent a difference of 0.1~dex in Li and [Fe/H], 
      and of 150~K in T$_{\rm{eff}}$}
\label{fig:comparisonotherworks}
\end{figure}

We have two, eight, and eighty two stars, respectively, in common with the samples of \citet{DeLaverny03},
\citet{Lebre_etal06}, and \citet{Lucketal07}.
We have nine stars in common with the Li sample of the AMBRE project \citep{GuiglionetalAMBRE16}; however the stellar effective temperatures are not provided in this study, and only two out of the nine common stars have had their [Fe/H] values published after being determined with AMBRE tools \citep{Mikolaitisetal17}. 
We have no stars in common with the samples of 
\citet{Brown_etal89}, \citet{Liu_etal14}, nor  the large GALAH survey \citep{GALAH2018}.

We show in Fig.~\ref{fig:comparisonotherworks} the comparison between our LTE Li results and the LTE Li values of the above-mentioned papers which include the same stars (see caption for details). 
We also compare the effective temperature values (since the Li abundance is more sensitive to this parameter than to others, and because T$_{\rm{eff}}$ is important in the comparison with stellar models, see \S~\ref{subsection:comp_models_obsLi}) and the [Fe/H] values. 

Our Li and T$_{\rm{eff}}$ determinations agree fairly well (within less than 0.2~dex and 150~K respectively) with those derived by \citet{Lucketal07} through their photometric pipeline. This was expected since we  used the same photometric processing (calibration using observed B-V and \citet{Houda} models for the determination effective temperature) and the same MARCS models for determining the Li abundances. 
Larger differences exist, however, between our analysis and the works of \citet{Lebre_etal06} and \citet{DeLaverny03}. 
These two studies relied on stellar parameters from the literature, which might have introduced some heterogeneity in their analysis; additionally, \citet{Lebre_etal06} report smaller S/N than we do here (> 50 for northern stars, > 80 for southern stars), and their Aurélie spectra for northern stars have a lower resolution (45\,000 instead of 70\,000), but their CES@CAT spectra have a higher resolution than ours (95\,000). The study of \citet{DeLaverny03} relies on the same observational material.

The bottom panel of Fig.~\ref{fig:comparisonotherworks} shows that our estimates of [Fe/H] tend to be lower than those of other authors by about 0.1 to 0.3~dex. 
The explanation for that difference certainly lies in the 
$2.0$\,km\,s$^{-1}$ micro-turbulence value we adopted, and which is larger by about 
$0.3$\,km\,s$^{-1}$ on average than the typical values determined by \citet{Lucketal07}.
Indeed, a larger micro-turbulent velocity results in a lower abundance, since the resulting 
synthetic lines are less saturated. The other two works involved in
Fig.~\ref{fig:comparisonotherworks} adopted $v_\mathrm{turb}=2.0$~km\,s$^{-1}$ but their
contribution amounts to only $10$ stars instead of $82$ for \citet{Lucketal07}.

\section{Comparison with stellar model predictions}
\label{section:comp_models_obs}

\begin{figure} 
	\centering
        \includegraphics[angle=0,width=9cm]{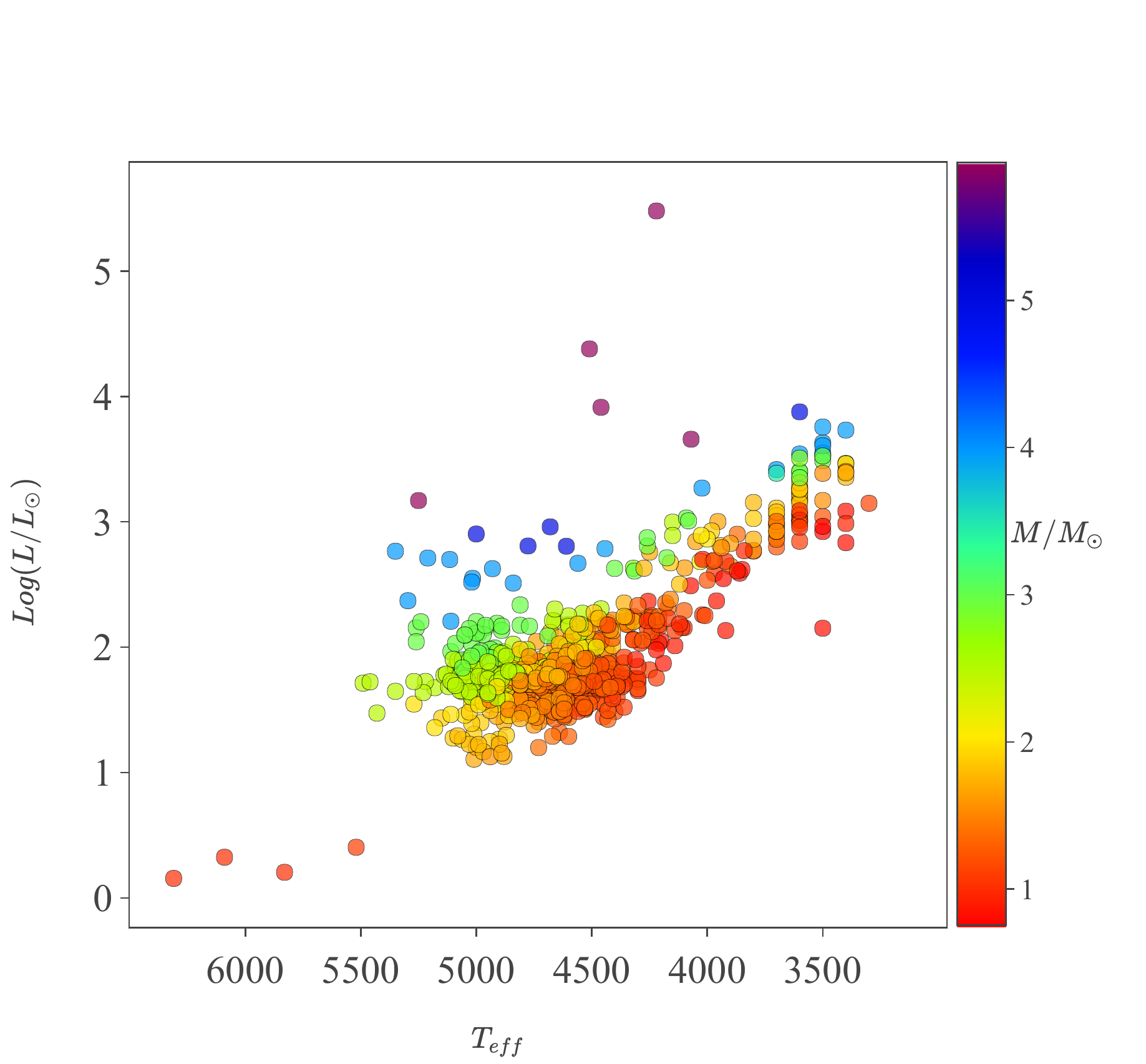} 
	  \caption{Current masses of our sample stars as derived in \S~\ref{subsection:massrangeevolutionstatus}}
\label{fig:massesoursample}
\end{figure}

\begin{figure*} 
	\centering
\includegraphics[angle=0,width=0.48\hsize,trim=1cm 5.8cm 1cm 3cm, clip=true]{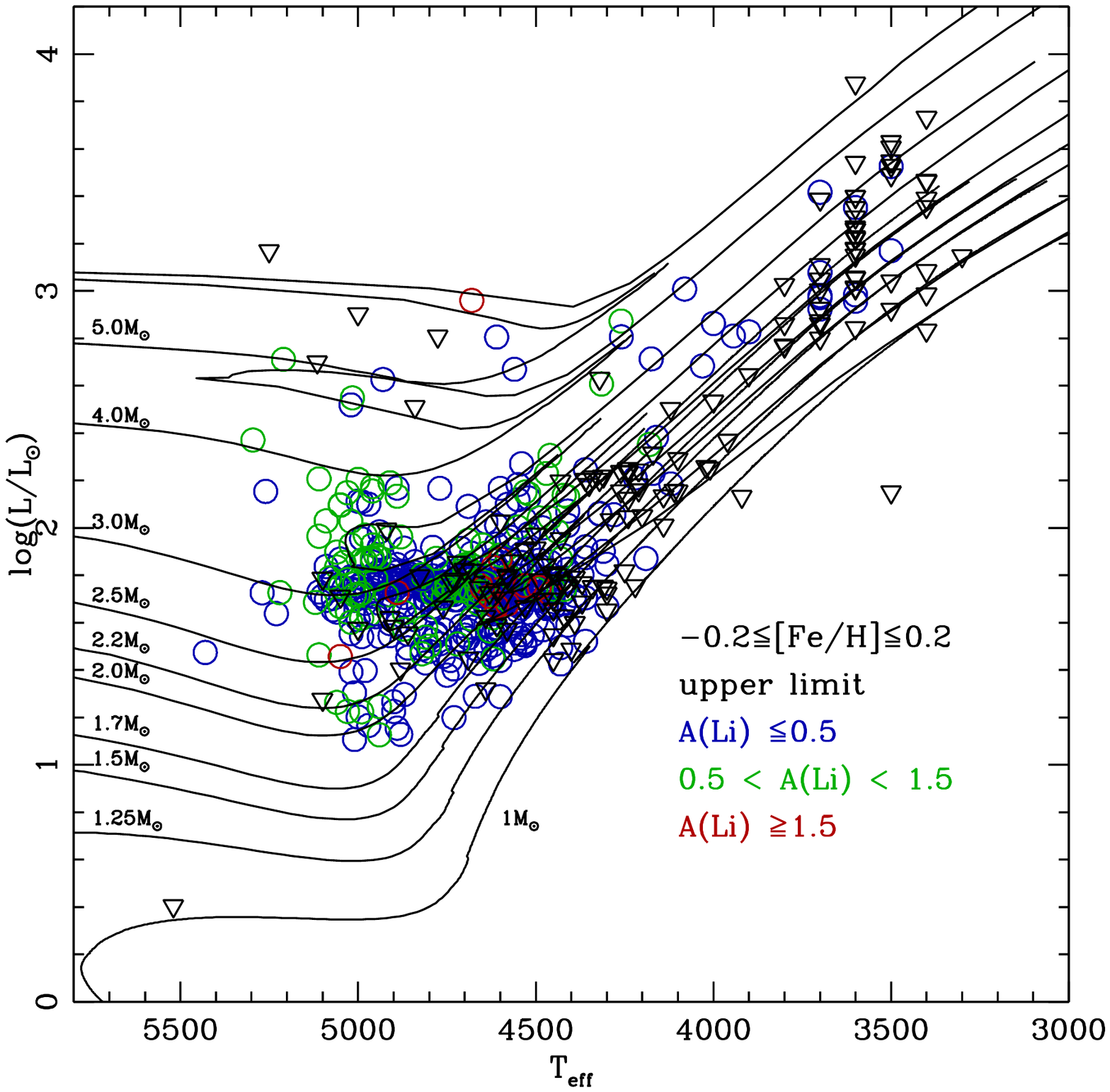}
\includegraphics[angle=0,width=0.48\hsize,trim=1cm 5.8cm 1cm 3cm, clip=true]{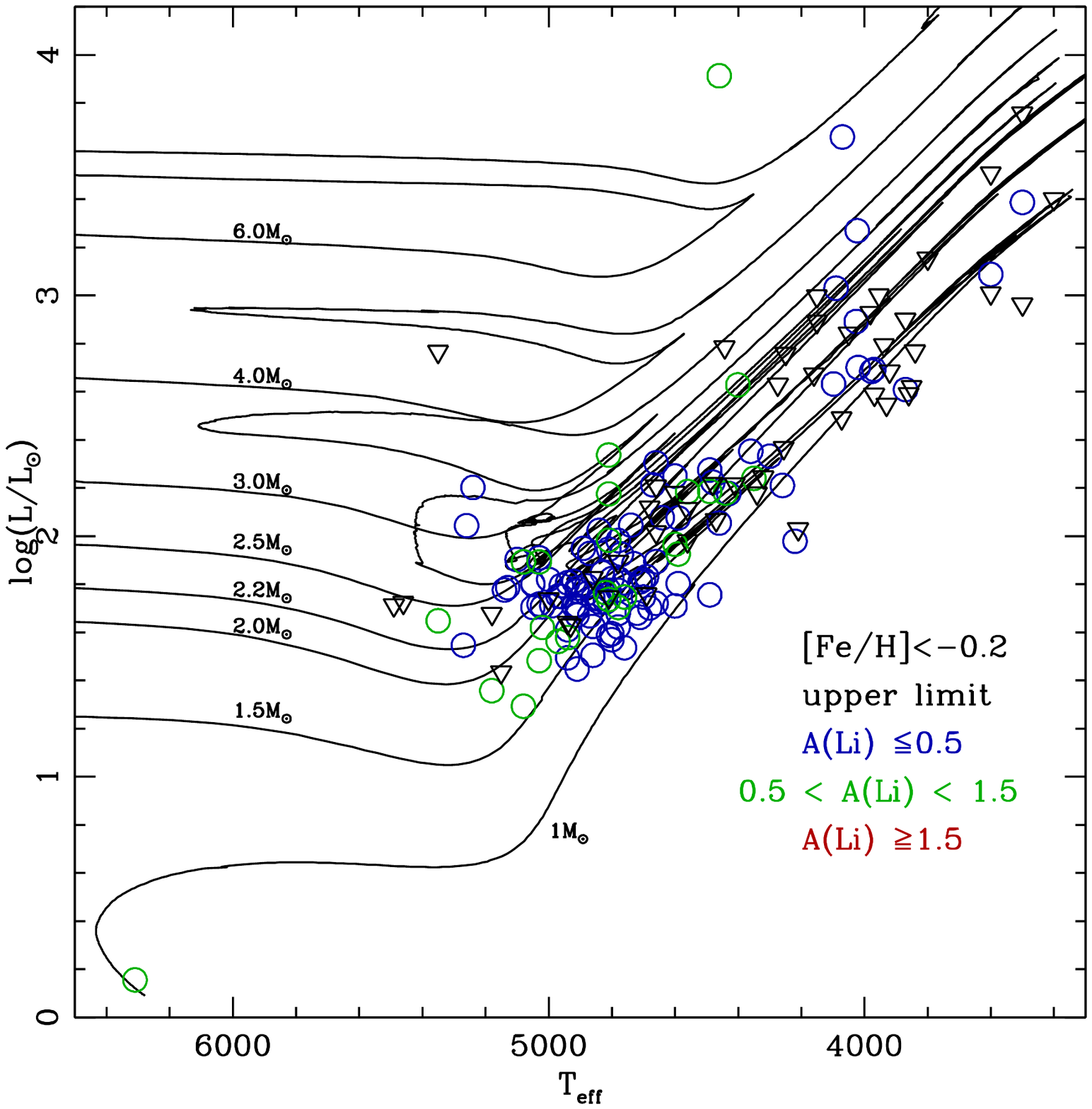}
	  \caption{Position of our sample stars in HRD with colour-labelled indication of Li (LTE) range as in Fig.~\ref{HR_LI_all}. 
	  The sample is split in two metallicity ranges ([Fe/H] above and below -0.2~dex, left and right respectively). 
	  The theoretical evolution tracks of the  standard models (no rotation) of \citet{Lagarde12a} are superimposed for stars with masses between 1 and 6~M$_{\odot}$ computed with [Fe/H]=0 and -0.56~dex (left and right, respectively); we removed the two outliers with errors on the parallax larger than 70~$\%$ 
      }
	\label{fig:HR_Li_tracks_Zsun}
\end{figure*}

\begin{figure*} [h]
	\centering
\includegraphics[angle=0,width=0.48\hsize,trim=1cm 5.5cm 1cm 3cm, clip=true]{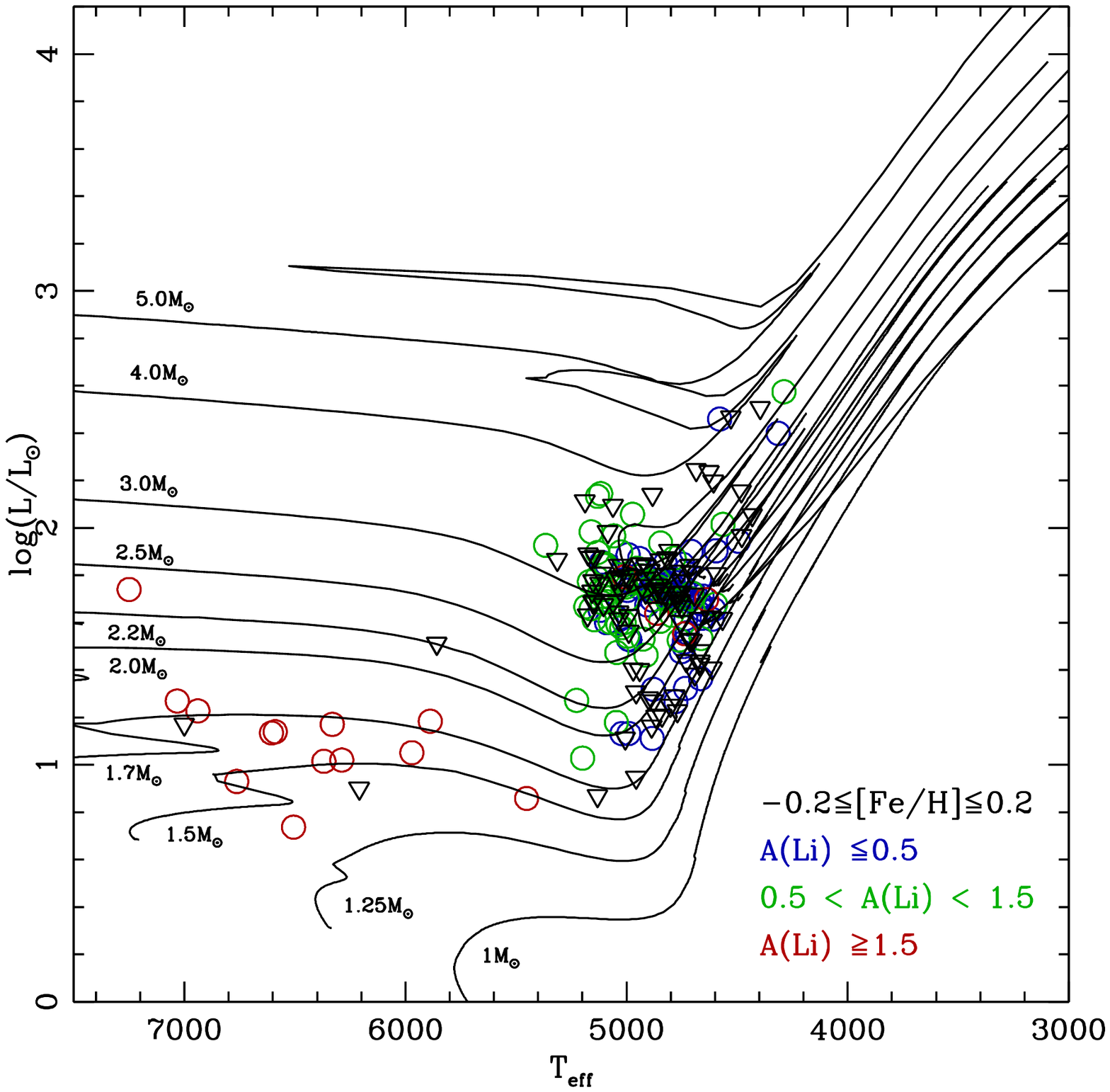}
\includegraphics[angle=0,width=0.48\hsize,trim=1cm 5.5cm 1cm 3cm, clip=true]{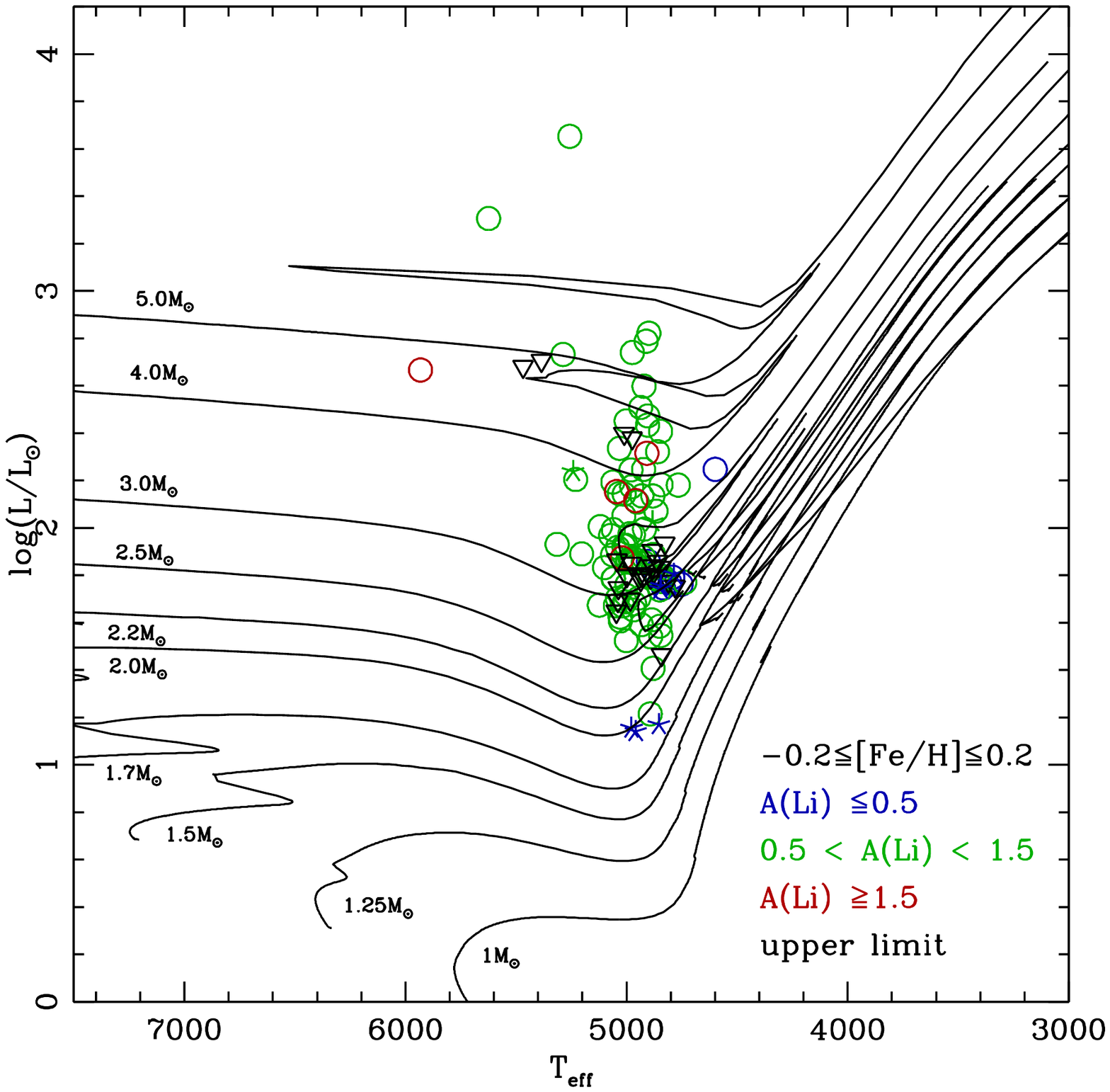}
	  \caption{ 
      Same as Fig.~\ref{fig:HR_Li_tracks_Zsun} but for samples observed by \citet[][LTE, left]{Lucketal07}  and \citet[][NLTE, right, with crosses indicating planet host stars]{Liu_etal14}. Here we show only their stars with $-0.2\leq$[Fe/H]$\leq+0.2$ that have \gdrtwo parallaxes (290 and 375 stars for \citet{Lucketal07} and \citet{Liu_etal14}, respectively). Evolution tracks are the standard models of \citet{Lagarde12a} computed with [Fe/H]$=0$  
      }
	\label{fig:HR_LH07}
\end{figure*}

\begin{figure*}[h]
	\centering
\includegraphics[angle=0,width=0.48\hsize,trim=1cm 5.5cm 1cm 3cm, clip=true]{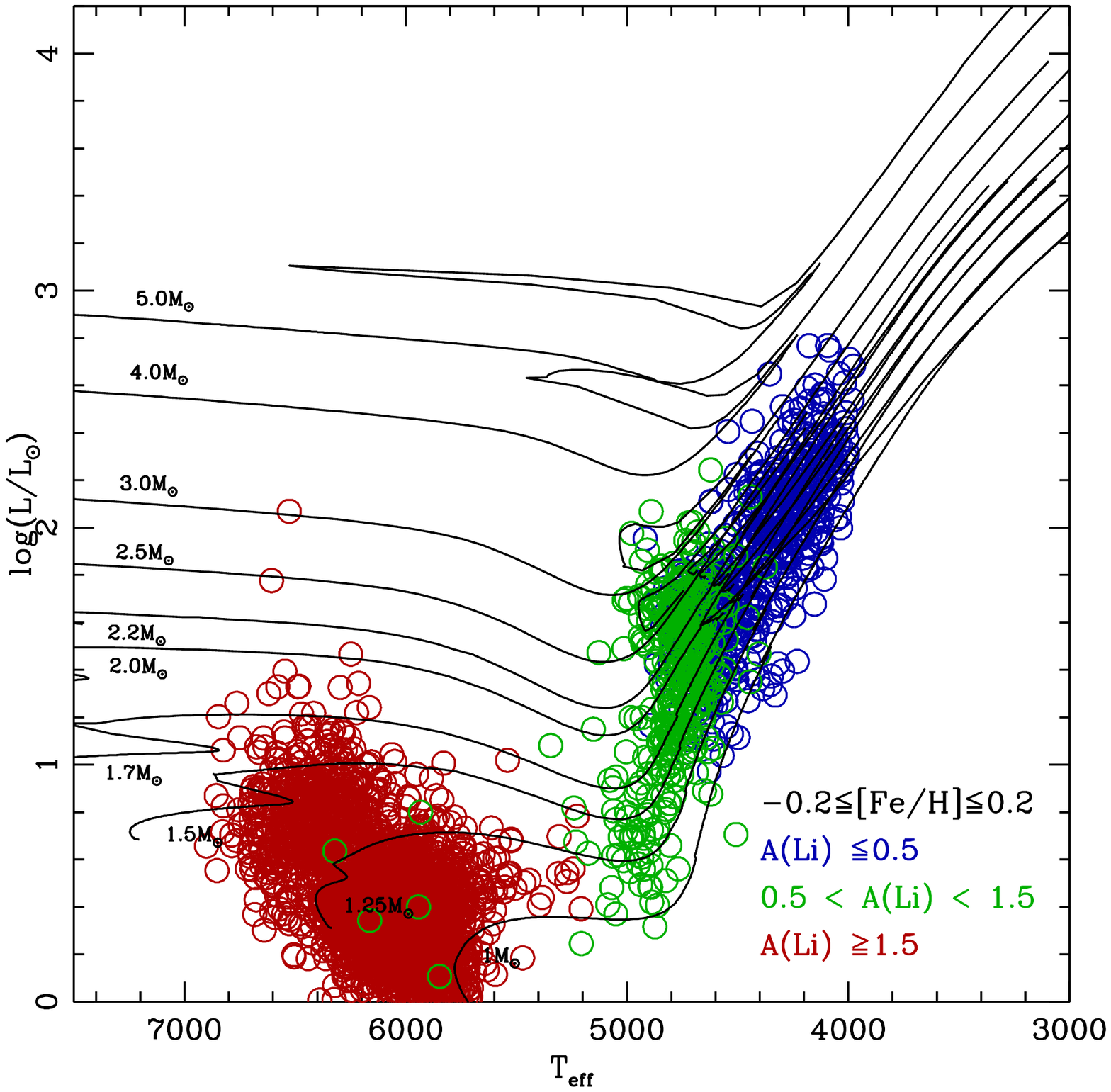}
\includegraphics[angle=0,width=0.48\hsize,trim=1cm 5.5cm 1cm 3cm, clip=true]{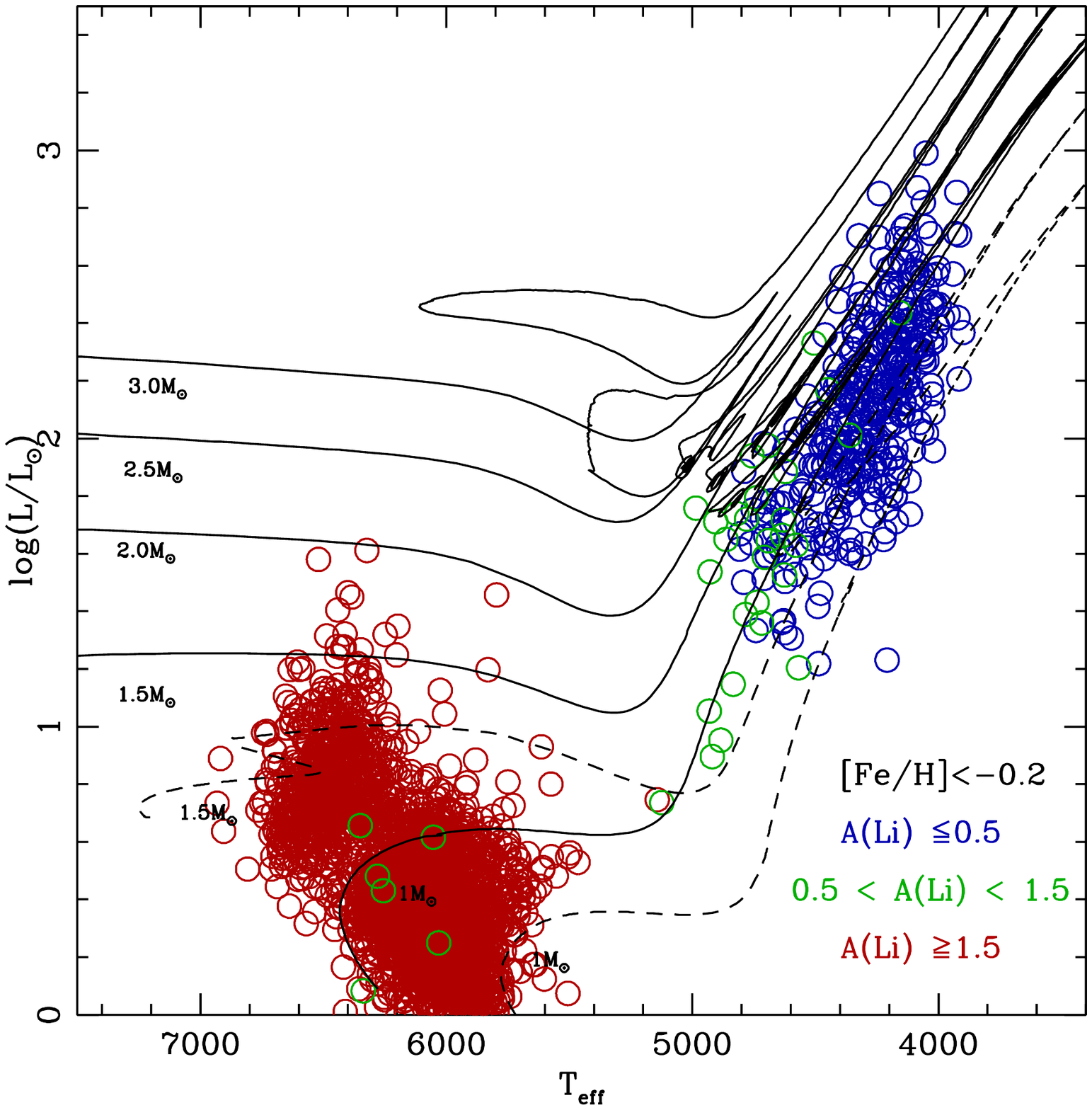}
	  \caption{
      Same as Fig.~\ref{fig:HR_Li_tracks_Zsun} for GALAH sub-samples with high accuracy or level of confidence in NLTE Li abundance (i.e. with Flag$_{Cannon}$=0 and Flag$_{Li} \leq 1$) and derived stellar parameters (\gdrtwo parallaxes with errors less than 20$\%$, and errors on T$_{\rm{eff}}$ less than 150~K), and with  $-0.2\leq$[Fe/H]$\leq+0.2$ and 
      [Fe/H]$<-0.2$  (left and right respectively). 
       Evolution tracks are the standard models of \citet{Lagarde12a} computed with [Fe/H]$=0$ (left), and [Fe/H]=0 and -0.56 (right, dashed and solid respectively)
      }
	\label{fig:HR_GALAH}
\end{figure*}

\subsection{Mass range and evolution status  of the sample stars}
\label{subsection:massrangeevolutionstatus}

In order to test model predictions for the Li behaviour along stellar evolution with respect to observations, we need to estimate the mass of individual stars.
For this, we applied the following method to our sample and to the samples of \citet[][]{Lucketal07}, \citet[][]{Liu_etal14}, and  \citet{GALAH2018}. 
We use a maximum-likelihood technique adapted from \citet{Valle_2014MCMC} that compares the effective temperature, luminosity, and [Fe/H] of individual stars to theoretical evolution tracks in the HRD. The interpolation is done between the standard models (without rotation) of \citet{Lagarde_etal17GBM,Lagardeetal19} that are available for six values of [Fe/H] (0.51, 0, -0.23, -0.54, -1.2, -2.14) for a range of masses between 0.6 and 6~M$_{\odot}$. The mass estimate accounts for the errors in effective temperature and luminosity given in Table~\ref{Observations_table} for our sample stars, as well as for the errors given in the original papers for the other samples; for all the samples we assume [Fe/H] error of 0.1~dex. The values we derived for the mass and the corresponding uncertainty are given in Table~\ref{Observations_table} (see also Fig.~\ref{fig:massesoursample}) for our sample stars and in Table~\ref{othersample} for the three other samples. For the GALAH survey, we derived the stellar mass for giant stars only (1734 stars). Because evolved stars lie in the region of the HRD where all the evolution tracks converge, and because of non-negligible error bars in effective temperature, luminosity, and metallicity, the masses we derive are relatively uncertain, as shown by the error bars we derive. None of the sample stars has an asteroseismic mass. Fortunately, very precise determination of the stellar mass of individual stars is not mandatory for this work, as we shall simply compare the Li abundance of stars in different domains of mass with the corresponding model predictions (\S~\ref{subsection:comp_models_obsLi}).

Our sample covers the largest mass and advanced evolution ranges (compare Fig.~\ref{fig:HR_Li_tracks_Zsun}, \ref{fig:HR_LH07}, and \ref{fig:HR_GALAH}). 
It contains stars with masses between $\sim 0.8$ and $6$~M$_{\odot}$, and evolution stages from the end of the Hertzsprung gap to the early-AGB, the majority lying in the region of the RGB bump and of the clump, and on the upper part of the RGB, as a result of our selection criteria.
On the other hand, the samples of \citet[][]{Lucketal07} and \citet[][]{Liu_etal14} focus on a more restricted mass range (between 1.5 and 4~M$_{\odot}$, and between 1.7 and 5~M$_{\odot}$ respectively; Fig.~\ref{fig:HR_LH07}). Additionally, most of their stars appear to lie around the clump, although \citet[][]{Lucketal07} also include LIMS (1.25 to 2~M$_{\odot}$) located between the MS turnoff and the RGB base.
Finally, GALAH sample (Fig.~\ref{fig:HR_GALAH}) contains many MS stars, as well as low-mass evolved stars with luminosities much lower than that of the RGB tip.

\subsection{Li behaviour - The impact of rotation-induced mixing and thermohaline mixing}
\label{subsection:comp_models_obsLi}

We compare the Li behaviour of our sample stars and of the samples of  \citet[][]{Lucketal07}, \citet[][]{Liu_etal14}, and GALAH \citep{GALAH2018} with the predictions from the grid of rotating stellar models of \citet{Lagarde12a}. These models take into account thermohaline mixing as described by \citet{charbonnel2007} and \citet{ChaLag10}, and rotation-induced processes following the formalism by \citet{zahn1992}
and \citet{maeder1998}. 
The initial rotation velocities assumed in the computations were chosen to lead to a good agreement with the observed range of rotation velocities in field and open cluster main sequence stars in the mass range of our sample (\citealt{Gaige1993} and \citealt{ZorecRoyer2012}; see \citealt{ChaLag10} and \citealt{Lagarde12a} for details). The corresponding evolution tracks are very similar to the standard tracks used in \S~\ref{subsection:massrangeevolutionstatus}
to estimate the mass and evolution stage of the sample stars, hence no correction is needed with respect to these two quantities. However, thermohaline and rotation-induced mixing processes are necessary to account for the Li behaviour along the explored evolution sequence over the entire mass range considered. This is shown in Fig.~\ref{fig:LivsTeff_modeles_Zsun}, \ref{fig:LivsTeff_modeles_lowZ}, \ref{fig:LH07_Li}, \ref{fig:Liu14_Li},  \ref{fig:GALAH_Li}, and \ref{fig:GALAH_Limetalpoor},  
where the model predictions are compared to the Li observations, with effective temperature as a proxy for the evolution  from the subgiant branch to the advanced phases (i.e., from high towards low temperatures). Standard model predictions are also shown for discussion. In each figure panel, we select stars in a given mass and metallicity range, and compare them with the relevant theoretical tracks that are shown from the zero age main sequence up to the AGB phase. 

\subsubsection{Our sample}
\label{Comp:oursample}

\begin{figure} 
	\centering
 \includegraphics[angle=0,width=9cm,trim=1cm 6.5cm 1cm 4cm, clip=true]{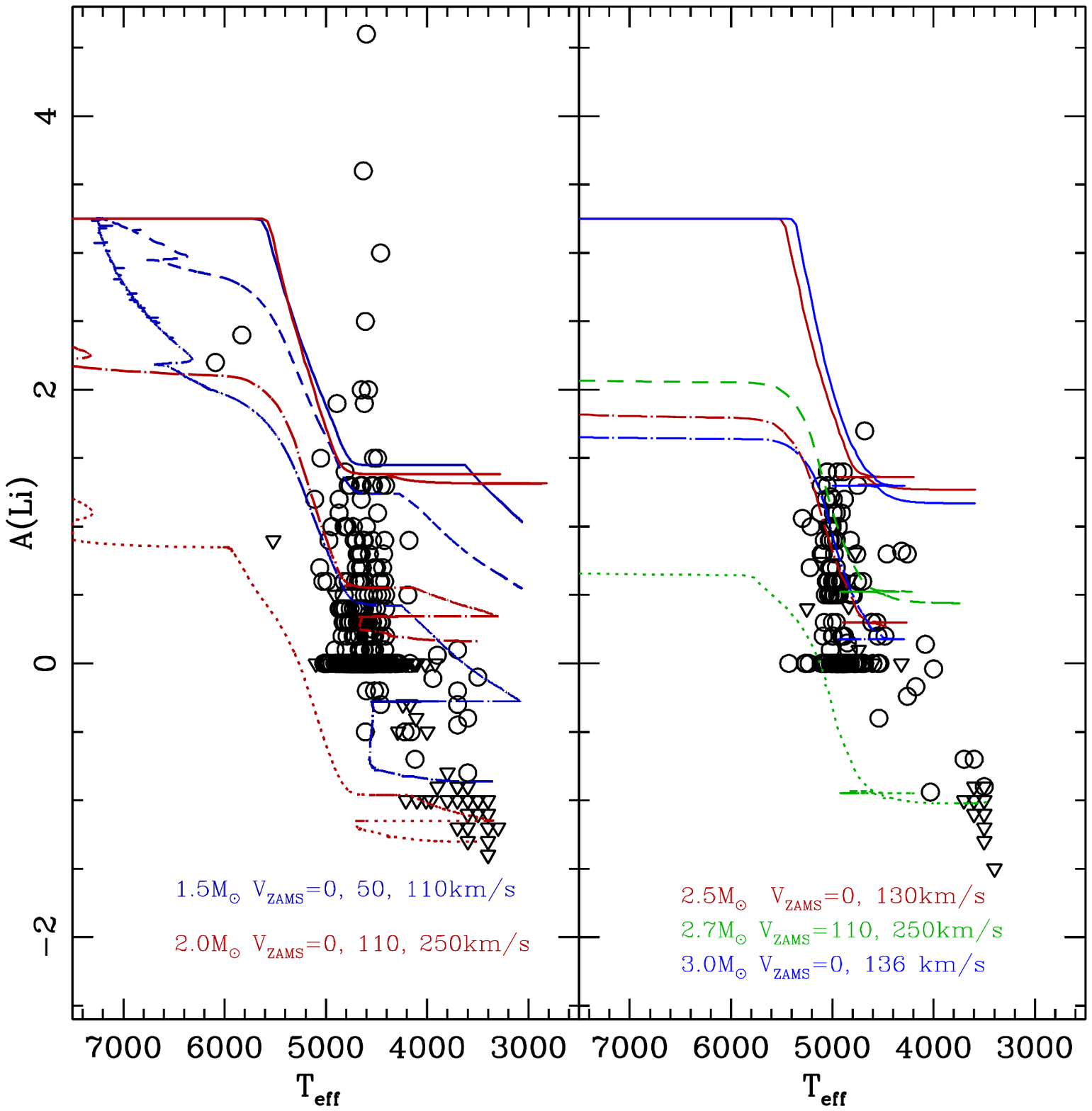}
	  \caption{Li data (LTE) for our solar-metallicity subsample (-0.2$\leq$[Fe/H]$\leq$+0.2; circles and triangles: Li determinations and upper limits) as a function of effective temperature as proxy for evolution (from high to cool values).
	  Tracks show the predictions for Li behaviour as the stars evolve from the subgiant branch to the RGB tip and later from the clump to AGB phase from the non-rotating models with thermohaline mixing (solid) and the models including rotation and thermohaline mixing (others) from \citet{Lagarde12a} 
      computed for [Fe/H]=0. At this metallicity, the initial Li abundance assumed in the models is 3.11~dex.
      In the left and right panels we plot respectively the stars with masses  $\leq$ and $>$ 2~M$_{\odot}$.       
      The initial masses of the models are colour-labeled, with information on the initial (ZAMS) rotation velocity 
        }
	\label{fig:LivsTeff_modeles_Zsun}
\end{figure}

\begin{figure} 
	\centering
   \includegraphics[angle=0,width=9cm,trim=1cm 6.5cm 1cm 4cm, clip=true]{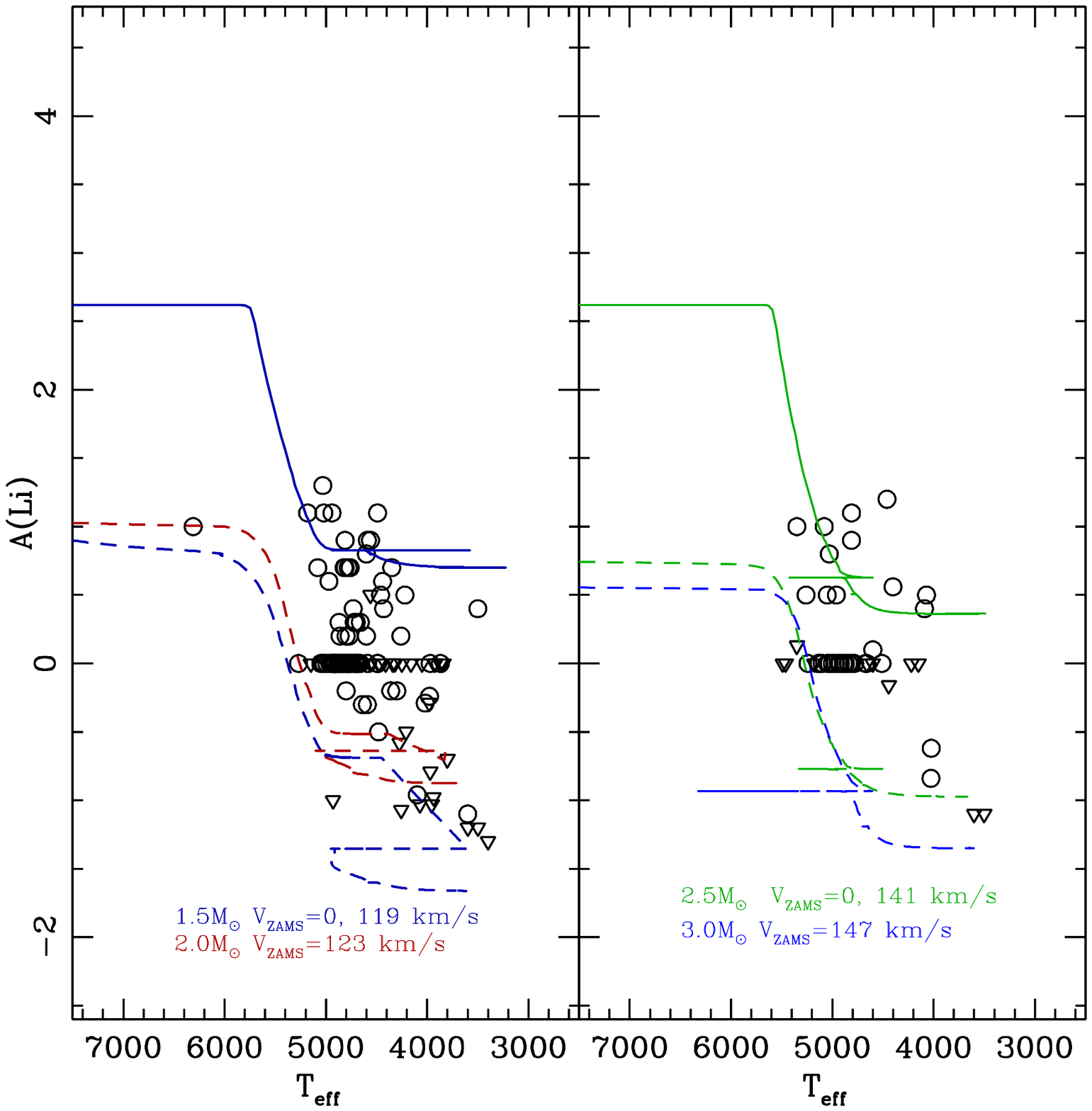}
	  \caption{Same as Fig.~\ref{fig:LivsTeff_modeles_Zsun} for our low-metallicity sub-sample ([Fe/H]$<$-0.2) and non-rotating models with thermohaline mixing (solid) and models including rotation and thermohaline mixing (others) of  \citet{Lagarde12a} computed for [Fe/H]=-0.56. At this metallicity, the initial Li abundance assumed in the models is 2.49~dex 
           }
	\label{fig:LivsTeff_modeles_lowZ}
\end{figure}

Let us focus first on our own sample, which we split in two metallicity ranges. 
In Fig.~\ref{fig:LivsTeff_modeles_Zsun} we gather the Li data for our sample stars with [Fe/H] between $+0.2$ and $-0.2$~dex (inclusive) and compare their behaviour with theoretical predictions for stellar models  computed with [Fe/H]$=0$~dex. In Fig.~\ref{fig:LivsTeff_modeles_lowZ} we show the Li data for the remaining stars with [Fe/H] strictly lower than $-0.2$~dex and use stellar models  computed with [Fe/H]$=-0.56$~dex. Note that for these two values of [Fe/H] the initial Li abundance assumed in the models is 3.11 and 2.49~dex respectively, to account for Galactic chemical evolution \citep[e.g.][]{Romano_etal99,Ryan_etal01,LambertReddy04,GuiglionetalAMBRE16,Fu_etal18}. 
In both cases we 
separate the stars with masses $\leq$ and $>$ 2~M$_{\odot}$ and compare them with models of relevant initial stellar masses.

We first note that the Li surface evolution for the standard (non-rotating, i.e. V$_{ZAMS}$=0) models is very similar for stars of different masses and metallicities. In this case, surface Li depletion is due solely to the first dredge-up, and it starts relatively late in the Hertzsprung gap, that is, for T$_{\rm{eff}}$ between $\sim$ 5200 and 5400~K.  
The predicted decrease in Li due to the first dredge-up is of the order of a factor 30 -- 60.
This depletion factor depends only on the maximum depth in mass reached by the convective envelope at the end of the dredge-up. It is independent of the initial Li value assumed in the models and it hardly depends on the stellar mass and metallicity. 
As a consequence, the minimum Li value obtained at the end of the first dredge-up episode is of the order of A(Li)$\sim$ 1.3 -- 1.6 for the models with [Fe/H]=0 and A(Li)$\sim$ 0.7 -- 1 for the models with [Fe/H]=-0.56. These standard predictions are in very good agreement with literature models (e.g. \citealt{Iben67,palacios2006,charbonnel2007,Mucciarellietal12}), but
at odds with the data. Clearly, the surface Li depletion observed starts earlier (i.e., for higher T$_{\rm{eff}}$) and is much more efficient than predicted by the standard models that account only for dilution during the first dredge-up. This confirms earlier findings \citep[e.g.][]{Alschuler75,Brown_etal89,Balachandran90,DoNascimento_etal00,palacios2003}.

On the other hand, the global behaviour and the dispersion of the Li abundances observed for the entire sample at all effective temperatures are very well reproduced by the rotating models. In this case, surface Li depletion starts much earlier (i.e., at a higher effective temperature at the entrance onto the Hertzsprung gap) than in the standard models. This is due to the enlargement of the Li-free region by rotation-induced mixing while the stars are still on the MS. This also leads to more substantial Li depletion at the end of the first dredge-up event, as already anticipated with the first generation of rotating models of \citet[][also calculated with STAREVOL but with a less sophisticated treatment of rotation]{palacios2003}. For a given stellar mass, the higher the initial rotating rate, the lower the Li abundance all along the evolution from the main sequence turnoff to the RGB and the AGB. Clearly, rotation-induced mixing  brings the models in very good agreement with the Li data over the entire mass range, with the dispersion being explained by the assumed spread in initial rotation rate. 
As mentioned in \S~\ref{subsection:abunddetermination}, NLTE corrections would amount to only 0.1 to 0.3~dex, with the highest values for the cooler stars, that is, in the effective temperature regime where we already have lots of upper limits for Li. Hence, the above conclusions are not statistically affected by the lack of corrections (see also  \S~\ref{subsection:LTEvsNLTE}).   

Additionally, for the low-mass range (below $\sim$ 2~M$_{\odot}$) 
thermohaline mixing occurs after the end of the first dredge-up event once the models reach the so-called RGB luminosity bump (which occurs at T$_{\rm{eff}}$ $\sim$ 4400 - 4300K, see Fig.~\ref{fig:LivsTeff_modeles_Zsun} and \ref{fig:LivsTeff_modeles_lowZ}, left panels). This leads to further Li depletion, as thermohaline mixing quickly transports Li from the convective envelope to hotter regions where it efficiently burns.
The corresponding transport coefficient is obtained with the prescription for the thermohaline diffusion coefficient ``\`a la Ulrich \& Kippenhahn" advocated by \citet{CZ07}. 
The excellent agreement 
between the Li data of low-mass stars that are more evolved than the RGB bump and our models provides strong support to our prescription for thermohaline mixing, despite existing tension with numerical simulations of this process \citep[e.g.][]{Denissenkov_10thermohalineRGB,Traxler_etal11,SenguptaGaraud18}. 
The same coefficient also accounts for the observed behaviour of the carbon isotopic ratio and of the abundances of C and N in the upper part of the red giant branch in Population I and II stars \citep{CZ07,ChaLag10,Lagarde12a}, and for the C and N abundances of giant stars investigated in the \gaia-ESO survey \citep{Lagardeetal19}. It also leads to a significant reduction of the production of $^3$He in low-mass stars, in better agreement with the behaviour of the evolution of this light element in the Galaxy \citep{CZ07,Lagardeetal11}.

In more massive stars (above $\sim$ 2~M$_{\odot}$; Figs.~\ref{fig:LivsTeff_modeles_Zsun} and \ref{fig:LivsTeff_modeles_lowZ}, right panels), thermohaline mixing is not expected to occur, because the stars do not pass through the RGB bump as they ignite central He burning earlier in non degenerate conditions.  For the upper mass range, the lowest Li abundance observed can simply be explained by higher initial rotation rates, as already discussed above. Again, the above conclusions for the upper part of the RGB (both for the low-mass and the more massive stars) would not be affected by NLTE corrections, since most of the cooler stars have only Li upper limits.

\subsubsection{Comparison with  \citet[][]{Lucketal07} and \citet[][]{Liu_etal14} samples}
\label{subsection:comparisonmodelsothersample}

For the samples of \citet[][]{Lucketal07} and \citet[][]{Liu_etal14} we considered only 
the stars that have both parallaxes and V magnitudes from \gdrtwo (290 out of 297 stars for \citealt[][]{Lucketal07}
and 375 out of 378 for \citealt[][]{Liu_etal14}). In this case, we used bolometric corrections from \citet{Alonso99} to compute the luminosities. The mass (and the error on the mass) of the sample stars were  determined as described in \S~\ref{subsection:massrangeevolutionstatus} and are listed in Table \ref{othersample}.

\begin{figure} 
	\centering
         \includegraphics[angle=0,width=0.98\hsize,trim=1cm 6.5cm 1cm 4cm, clip=true]{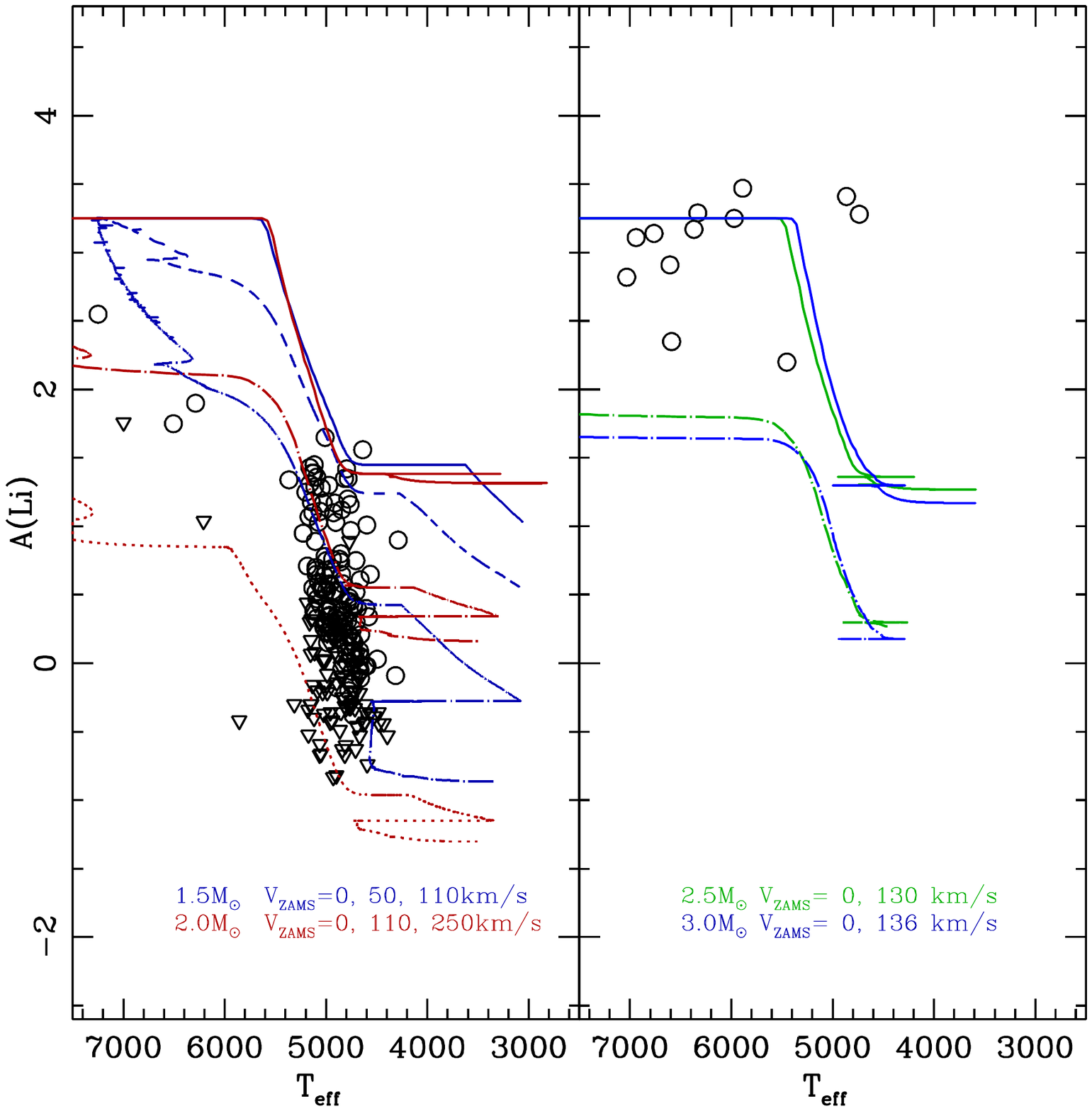}
        \caption{Same as Fig.~\ref{fig:LivsTeff_modeles_Zsun} (LTE abundances) for  
        samples observed by \citet{Lucketal07}.
        We show only their stars with -0.2$\leq$[Fe/H]$\leq$0.2 and compare with solar metallicity models 
        including rotation and thermohaline mixing from \citet{Lagarde12a} computed for [Fe/H]=0
      }
	\label{fig:LH07_Li}
\end{figure}

\begin{figure} 
	\centering
        \includegraphics[angle=0,width=0.98\hsize,trim=1cm 6.5cm 1cm 4cm, clip=true]{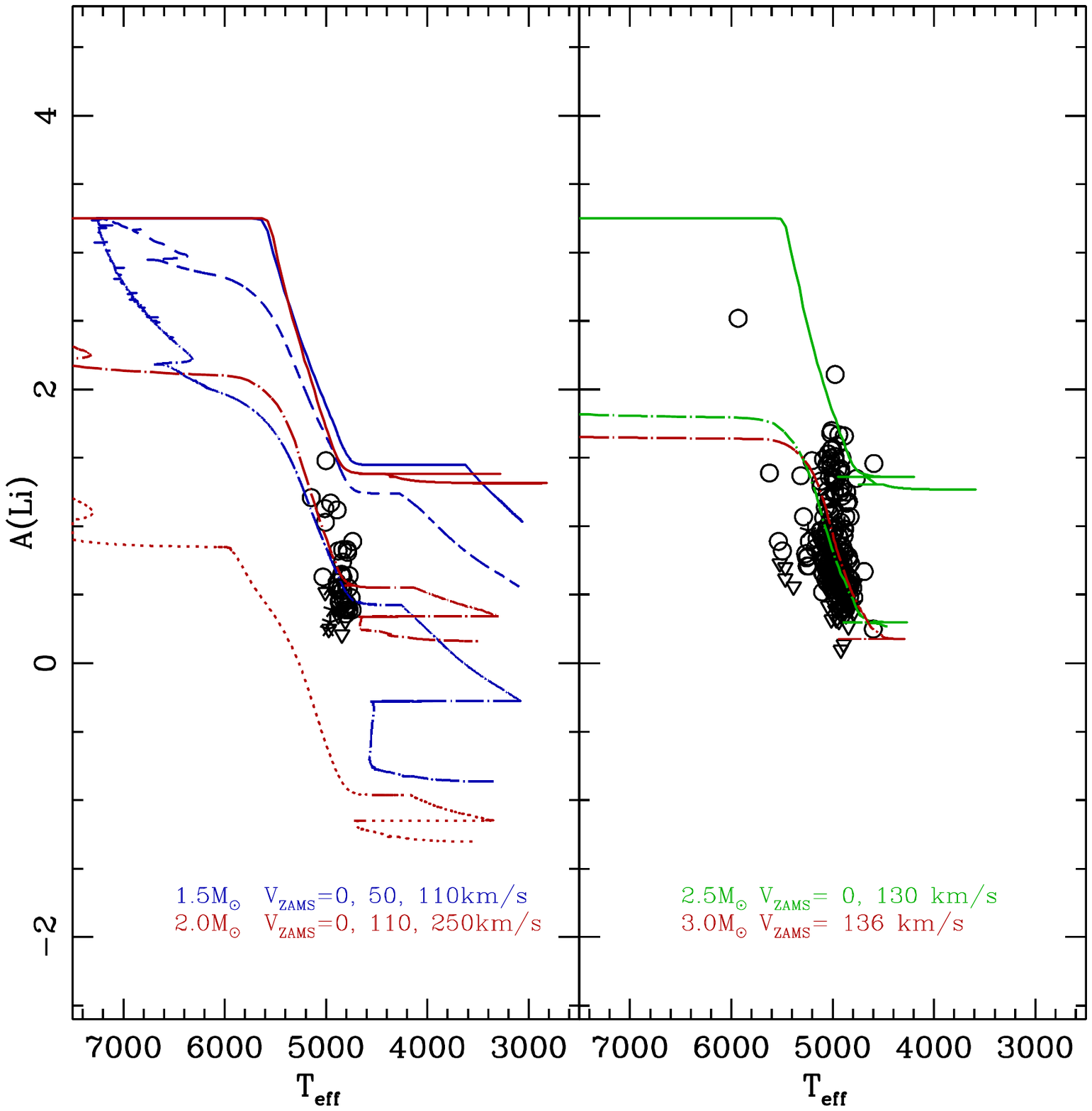}
        \caption{Same as Fig.~\ref{fig:LivsTeff_modeles_Zsun} for  
        samples observed by  \citet{Liu_etal14}; in this case, Li data is NLTE.
        We show only their stars with -0.2$\leq$[Fe/H]$\leq$0.2 and compare with solar metallicity models 
        including rotation and thermohaline mixing from \citet{Lagarde12a} computed for [Fe/H]=0
      }
	\label{fig:Liu14_Li}
\end{figure}

The predictions of the same rotating models at solar metallicity are compared in Fig.~\ref{fig:LH07_Li} and Fig.~\ref{fig:Liu14_Li} to the observations of the stars with [Fe/H] between +0.2 and -0.2 (inclusive) of the samples of \citet[][]{Lucketal07} and \citet[][]{Liu_etal14}. 
In the first case, the Li abundances are LTE, while in the second case we show the NLTE abundances provided by the authors. 
We find similarly good agreement for both, better than in \S~\ref{Comp:oursample}. 
Although we unfortunately have no stars in common with \citet[][]{Liu_etal14}, we see that the inclusion of NLTE corrections does not affect our global  conclusions. 
In addition to the conclusions derived with our own sample, the Li data of the less evolved stars of \citet[][]{Lucketal07} confirm the importance of rotation-induced mixing for the evolution of the surface Li abundance already at the MS turnoff and along the Hertzsprung gap.

\begin{figure} 
	\centering
        \includegraphics[angle=0,width=0.98\hsize,trim=1cm 6cm 1cm 4cm, clip=true]{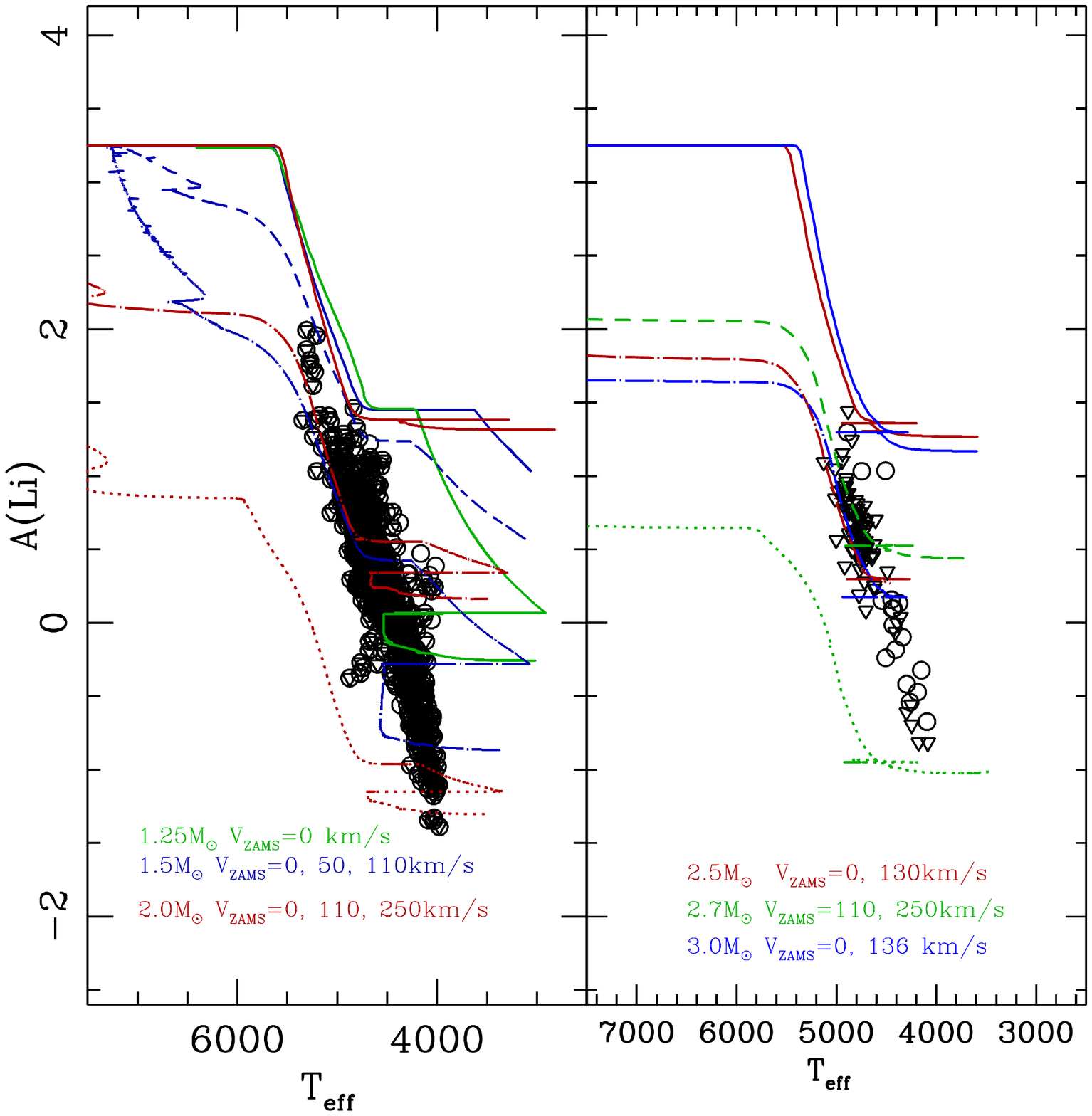}
        \caption{Same as Fig.~\ref{fig:LivsTeff_modeles_Zsun} for GALAH
        subsample of giant stars with $-0.2\leq$[Fe/H]$\leq 0.2$
        compared to predictions of models including rotation and thermohaline mixing computed with [Fe/H]$=0$. In this case, Li data is NLTE
      }
	\label{fig:GALAH_Li}
\end{figure}

\begin{figure} 
	\centering
        \includegraphics[angle=0,width=0.98\hsize,trim=1cm 6cm 1cm 4cm, clip=true]{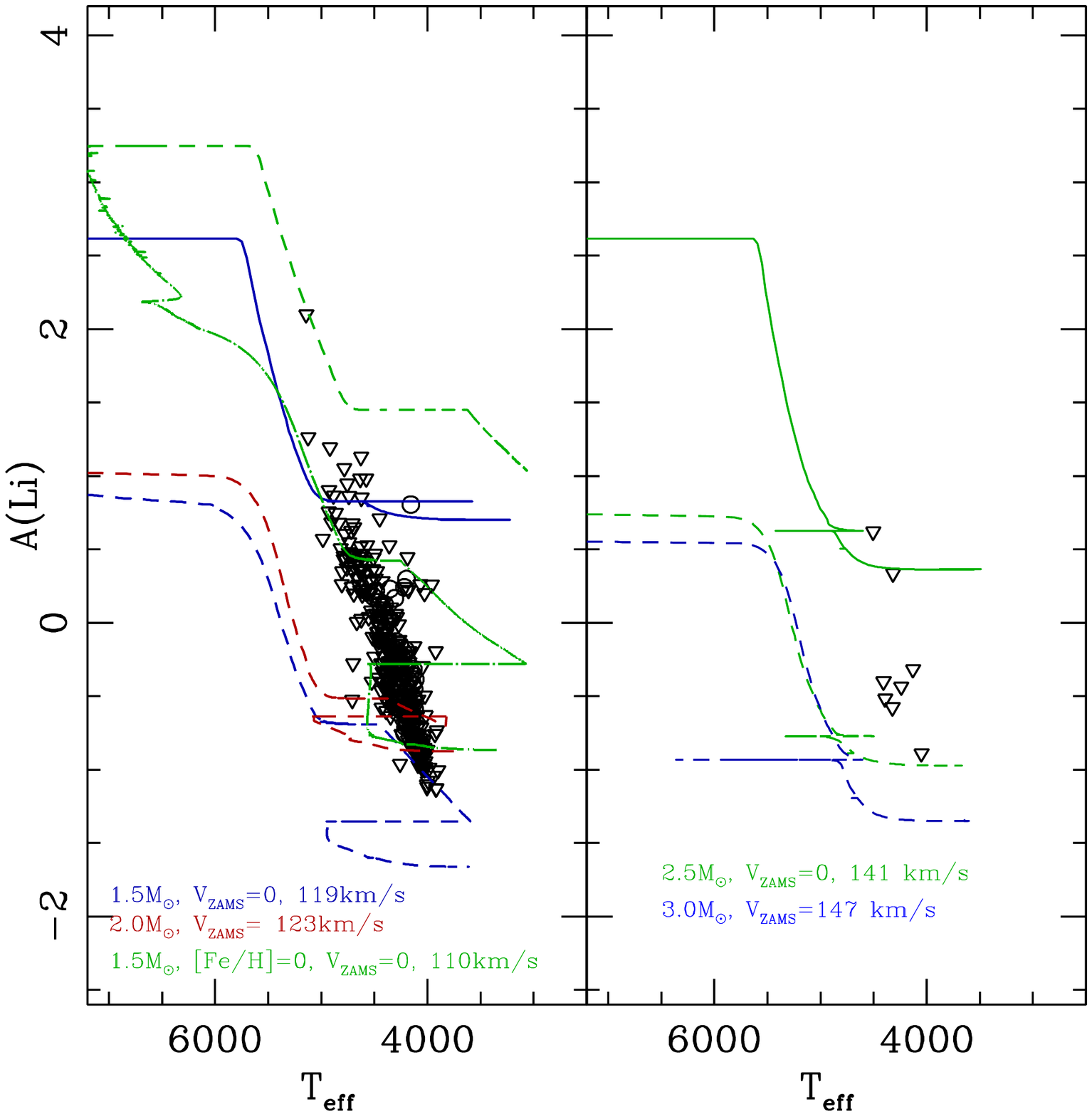}
        \caption{Same as Fig.~\ref{fig:GALAH_Li} for the GALAH
        sub-sample of giant stars with [Fe/H]$<$-0.2 compared to predictions of models including rotation and thermohaline mixing computed with [Fe/H]=-0.5 (and 0 green lines on the left panel only)       }
	\label{fig:GALAH_Limetalpoor}
\end{figure}

\begin{figure}[h]
\centering
        \includegraphics[angle=0,width=0.98\hsize,clip=true]{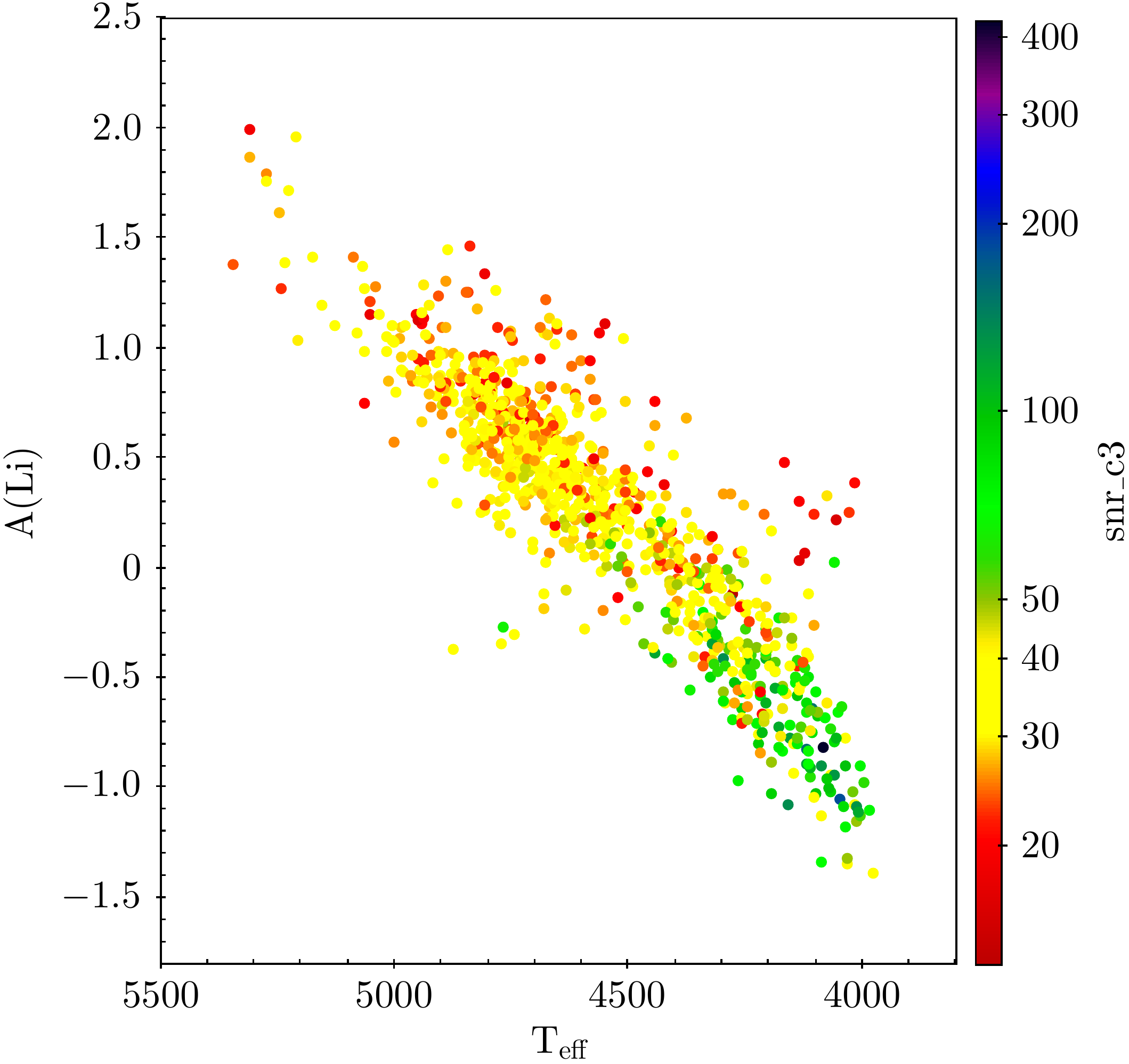}
\caption{Colour-coded S/N from \citet{GALAH2018} for the GALAH sub-sample selected with the criteria described in the text}
\label{fig:HR_GALAH_SNR3}
\end{figure}

\subsubsection{Comparison to GALAH sample with \gdrtwo parallaxes}
\label{subsection:GALAH}
In DR2 of the GALAH survey (Galactic Archaeology with HERMES; \citealt{GALAH2018}), NLTE Li abundances were delivered for 342682 stars, most of them having \gdrtwo parallaxes.
For comparison with stellar evolution models predictions, we considered only GALAH stars with high accuracy or level of confidence in derived stellar parameters and Li abundance (i.e., with Flag$_{Cannon}$=0 and Flag$_{Li} \leq 1$), \gdrtwo parallaxes with errors less than 20$\%$, and errors on T$_{\rm{eff}}$ less than 150~K. These selection criteria yielded a total of 
7829 stars (MS and giants). We used bolometric corrections from \citet{Alonso99} to compute the luminosities. The mass (and the error on the mass) of the sample stars was determined as described in \S~\ref{subsection:massrangeevolutionstatus}. The positions of the stars in the HRD are compared with the standard evolution tracks of \citet{Lagarde12a} in Fig.~\ref{fig:HR_GALAH} (we show only the stars with [Fe/H]$\leq$+0.2). 

For the comparison with model predictions for Li, we focussed on evolved stars (T$_{\rm{eff}}$ lower than 5500K, that is, 1734 stars in total). To test for metallicity effects, we select two subsamples 
with [Fe/H] between $-0.2$ to $+0.2$ (inclusive; 996 stars) on the one hand and [Fe/H] below $-0.2$ (391 stars) on the other hand. 
The behaviour of the NLTE Li abundance along evolution is shown in Figs.~\ref{fig:GALAH_Li} and \ref{fig:GALAH_Limetalpoor}, with relevant models for different mass ranges.

A striking relation becomes apparent between A(Li) and T$_{\rm{eff}}$, which is not observed in any of the other samples. Thus, we looked for the differences in the data that could explain this feature. We recall that our sample and the GALAH sample have no star in common, and that their spectral resolution (R$\sim$ 28'000) is lower than ours (see \S~\ref{observations}). We checked the SNR given by \citet{GALAH2018} and, in particular, their snr\_c3 which corresponds to the HERMES red channel covering the relevant wavelength range for Li. As can be seen in Fig.~\ref{fig:HR_GALAH_SNR3}, S/N is of the order of 20 -- 30 for the large majority of the GALAH sub-sample stars. Only nine stars have high S/N $\sim$ 150 as in our own data, out of which only four have a metallicity close to solar; these are also the stars with the lowest derived Li abundance and with the lowest effective temperature. If we keep the GALAH stars with the highest S/N, the sample obviously becomes too small to carry any statistical significance. On the other hand, the A(Li) and T$_{\rm{eff}}$ relation relies on very low S/N data, with the few high S/N stars having very low Li. This apparent relation is thus highly questionable in the context of the criteria that form the basis our own work, in particular, the analysis of high S/N, high resolution spectra.

\subsubsection{LTE versus NLTE}
\label{subsection:LTEvsNLTE}
Correcting the Li abundances for NLTE effects would raise the representative points in 
Figs.~\ref{fig:LivsTeff_modeles_Zsun},\,\ref{fig:LivsTeff_modeles_lowZ}, and \,\ref{fig:LH07_Li} by about $0.1$ to $0.33$~dex along the y-axis.
This would not change the above conclusions obtained using LTE Li abundances, especially as they are based on a whole distribution of observed Li abundances, many of which are upper limits.
A comparison between Figs.~\ref{fig:LivsTeff_modeles_Zsun},\,\ref{fig:LivsTeff_modeles_lowZ},\,\ref{fig:LH07_Li} and 
Fig.\ref{fig:Liu14_Li},\,\ref{fig:GALAH_Li},\,\ref{fig:GALAH_Limetalpoor} demonstrates that our conclusions di indeed remain valid for studies providing NLTE Li abundances as well as
for those providing LTE Li abundances. Thus, NLTE corrections do not appear crucial in our
context; this could be expected since the NLTE corrections amount to a few tenths of a dex,
while the effects of mixing reach about $2$~dex.

\subsection{Li-rich stars}
\label{subsection:Li-richstars}

\begin{figure} 
	\centering
        \includegraphics
        [angle=0,width=0.9\hsize]{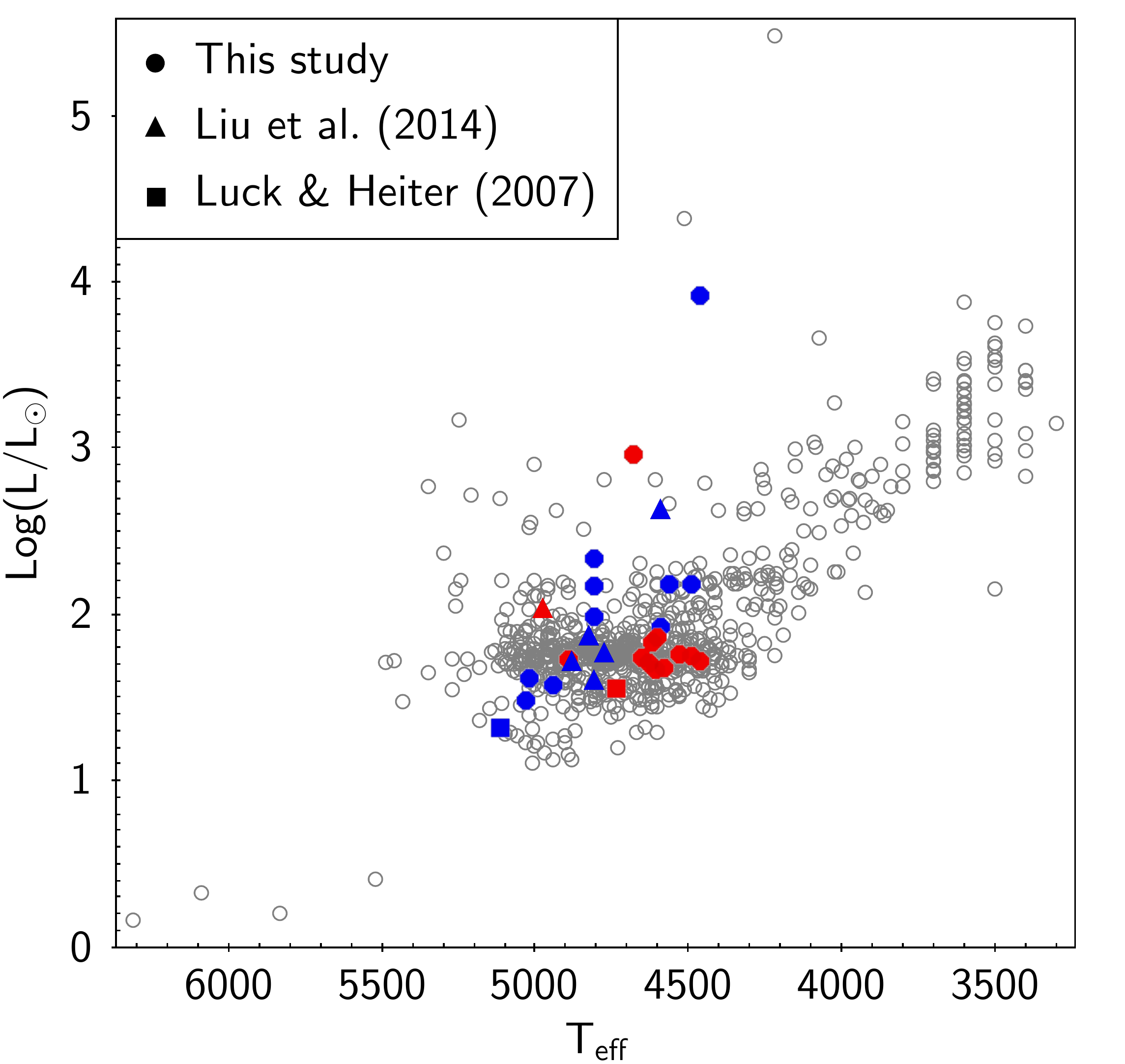}
        \caption{Position in HRD of the LiRG identified with our adopted criterion. Red and blue stars are for Li values higher or lower than 1.5~dex, respectively.
      }
	\label{fig:Lirich}
\end{figure}

\subsubsection{Li-rich giants: where do we stand?}
Li-rich giants (LiRG) are usually defined as evolved stars with A(Li) higher than 1.5~dex, which is the mean post dredge-up value predicted by standard evolution for Population I stars (see \S~\ref{Comp:oursample}). They are distinct from AGB stars producing Li through the \citet{CameronFowler71} mechanism during the thermal pulse phase \citep[e.g.][]{SmithLambert1989,ForestiniCharbonnel97,Dominguez_etal2004}.
Since the discovery of the first LiRG \citep{WallersteinSneden1982},
several surveys have reported the existence of such objects which account for $\sim$ 1 - 2 $\%$ of the studied populations \citep[][]{Brown_etal89,CharbonnelBalachandran2000,ReddyLambert2005,Lebre_etal06,Gonzalez_etal09,Kumaretal2011,Monaco_etal2011,Lyubimkov_etal12,Casey_etal16,Smiljanic_etal18,Yan_etal18,Singh_etal19,DeepakReddy2019}. 

Some studies use Hipparcos or \gaia data to determine the evolutionary status of field LiRG and show that these objects tend to accumulate close to the RGB bump, the clump, and the early-AGB (e.g. \citealt{CharbonnelBalachandran2000,Kumaretal2011,Smiljanic_etal18,DeepakReddy2019}), which is in agreement with open cluster studies 
\citep[e.g.][]{DelgadoMenaetal16}. 
Other works report, however, that LiRG can be randomly located in the HRD \citep{Jasniewicz_etal1999,Smith_etal1999,Monaco_etal2011,Lebzelter_etal2012,MartellShetrone2013,Casey_etal16}. 
The distinction is crucial to understanding the processes that may provide an explanation for the the phenomenon, such as fresh Li production by internal mixing processes \citep{SackmannBoothroyd1999,palaciosetal2001,Guandalini_etal2009,Strassmeier_etal2015,Cassisi_etal2016}, prompt mass loss events \citep{DeLaReza_etal1996,DeLaReza_etal1997}, Li accretion during engulfment of planets or planetesimals \citep{Alexander1967,SiessLivio1999,Carlberg_etal2010,AguileraGomezetal2016a,AguileraGomezetal2016b,DelgadoMenaetal16}, tidal interactions between binary stars  \citep{Casey_etal19}, or a combination of these mechanisms \citep{DenissenkovWeiss2000,DenissenkovHerwig2004}. 
However, since the evolution tracks of evolved stars all converge to the same area of the CMD, the definitive determination of the actual evolution status of LiRG requires  asteroseismology to probe their internal structure and disentangle RGB from clump stars. 
As of today, very few LiRG have been observed with CoRoT and {\it{Kepler}}. 
The majority seems 
to be located in the core-He burning clump  \citep{SilvaAguirre_etal_2014,BharatKumar_etal2018,Casey_etal16,Smiljanic_etal18,Singh_etal19}, with 
the others being at the RGB bump or higher on the first ascent giant branch \citep{Jofre_etal2015,Casey_etal19}.  
In a recent study using LAMOST spectra to derive both the Li abundance and asteroseismic classification, \citet{Casey_etal19} showed that $\sim 80 \%$ of their large sample of low-mass LiRG (2330 objects) probably have helium burning cores. They find that LiRG are more frequent at higher metallicity.

\subsubsection{Li-rich giants in the samples considered in the present study}
\label{LRGlist}
As mentioned above, the lower Li abundance for a LiRG is usually taken equal to 1.5~dex.
This refers to a standard first dredge-up dilution factor of $\sim$ 60, for a star with an initial cosmic abundance of 3.3~dex. 
Although this is in principle relevant for Pop~I giant stars that have already undergone the first dredge-up, this is not an appropriate criterion for less evolved stars that have not yet undergone full Li dilution. Additionally, more metal-poor stars were probably born with a lower initial Li abundance, with a value of $\sim$ 2.5~dex for stars with $\sim$ 3 times fewer metals than the Sun (discussion and references in \S~\ref{Comp:oursample}).  
Finally, uncertainties in the actual Li abundances should be taken into account, including those related to NLTE corrections. 

It is for these reasons that we identify as LiRG those stars that have LTE Li abundance 
and 0.2~dex higher than the Li value predicted by the standard models at the corresponding effective temperature and for the relevant value of [Fe/H] (see Fig.~\ref{fig:LivsTeff_modeles_Zsun} and related figures). We chose this value of 0.2~dex as it corresponds to the typical uncertainty for the Li abundance. 
We provide information on the LiRGs of our own sample and that of  \citet{Lucketal07} and \citet{Liu_etal14} 
in Tables~\ref{Table:Li-rich_oursample}, \ref{Table:Li-rich_othersamples} 
respectively. The position of all the LiRG we identified is shown in the HRD in Fig.~\ref{fig:Lirich}. 

In our own sample (Fig.~\ref{fig:Lirich} and Table~\ref{Table:Li-rich_oursample}) we identify seven LiRG that pass the ``classical" criterion (A(Li)$\geq 1.5$~dex). 
This corresponds to $0.9\%$ of the stars we observed, which a typical percentage for LiRG found in the surveys of the literature. 
With our new criterion, we obtain twelve additional LiRG. The total nineteen stars correspond to  $2.2\%$ of our sample.
Only two LiRG have Li higher than the cosmic value of 3.3~dex, one has a Li abundance of 3.0~dex; all the others have A(Li) $\leq 2.5$~dex.
All of them lie close to the RGB bump and the He-burning clump, except for two objects that are much brighter and more massive than the rest of the LiRG subsample, and which have relatively low [Fe/H]. The position of the low-mass LiRG is, thus, in agreement with the previous findings. Without asteroseismology however, no definitive conclusion can be obtained on their actual evolution state.

Upon applying our criterion, we find two and six LiRG
in the samples of \citet{Liu_etal14} and  \citet{Lucketal07}, respectively. This corresponds to 0.5 and 2~$\%$ of the total number of red giant stars considered from the respective surveys. 
We find no LiRG among our GALAH  subsample while 
\citet{DeepakReddy2019} 
found 0.64~$\%$ LiRG
and 0.03~$\%$ super Li-rich giant (A(Li)$>$3.2) 
among a larger GALAH subsample selected using a less restrictive Li abundance flag than the one we adopted (\S~\ref{subsection:GALAH}). All the LiRG we identify in the literature samples lie close to the RGB bump or the He-burning clump 
except for one star that is both brighter and more massive than the bulk LiRG sample. Again, no definitive conclusion can be obtained on their actual evolution state without asteroseismic constraints.

\subsubsection{In terms of planets}
\label{subsection:planets}
The entire LiRG sample has been cross-matched with the current sample of 112 published substellar companions around giant stars, a list kept up-to-date by Sabine Reffert’s Heidelberg 
team\footnote{https://exoplanetarchive.ipac.caltech.edu} 
and no match was found. In order to avoid any error in the determination of the host evolutionary stage, we extended our search to the current sample of 4003 published confirmed exoplanets\footnote{http://www.lsw.uni-heidelberg.de/users/sreffert/giantplanets/giantplanets.php}, once again leading to no match.

\section{Conclusions}
\label{section:conclusions}
Li is a fragile element that can be used as a tracer of transport processes in stellar interiors along the evolution of LIMS. Its photospheric abundance is known to vary extensively with initial stellar mass and evolution stage. Here we test stellar evolution models which predict stronger Li depletion for higher initial rotation in the advanced phases of the evolution as well as a strong impact of thermohaline mixing on low-mass, bright RGB stars. For this purpose, we derived Li abundances in a volume-limited sample of field giant stars with precise \gaia parallaxes and we also used literature samples with Hipparcos or \gaia measurements. The model-data comparison confirms the impact of rotation-induced mixing and provides excellent support to our prescription for the efficiency of thermohaline mixing. Finally, we find only $0.8\%$ of LiRG in our sample when using the classical definition used in the literature, and $2.2\%$ when choosing a definition that accounts for the actual evolution stage of the stars with respect to the end of the first dredge-up. 
We can conclude that \gaia puts a new spin on how the mixing processes in stars are understood. However, asteroseismology is required to definitively pinpoint the actual evolution status of the LiRG. 

\begin{landscape}
\begin{table}
\caption{Informations on giants stars in our sample. $flag_{Li}$ indicates the determination and the upper limit of Li abundances (0 and -1, respectively). $flag_{L}$ indicates the method used to compute the luminosity log(L/L$_\odot$) (0 : $\pi_{DR2}$+V$_{Evans 18}$ ; 1 : $\pi_{vl07}$+V$_{TYCHO}$ ; 2 : $\pi_{vl07}$+V$_{Hipp97}$ ; 0 : $\pi_{DR2}$+V$_{TYCHO}$). e$_{M [-1\sigma]}$ and e$_{M [+1\sigma]}$ correspond to the error on the mass determination at -1$\sigma$ and +1$\sigma$, respectively.
}        
\begin{center}
\scalebox{0.8}{
\label{observations_table}      
\centering                       
\begin{tabular}{|r|r|r|r|r|r|r|r|r|r|r|r|r|r|r|r|r|r|r|r|r|r|}       
\hline               
HD & RA & DEC &  T$_{eff}$ & e(T$_{eff}$) & BC & 
[Fe/H] & A(Li) & flag$_{Li}$ & $\pi_{vL07}$ & e$(\pi_{vL07})$ & L$_{Hipp VL}$ & e$(L_{Hipp VL})$& $\pi_{\gaia}$ & e$(\pi_{\gaia})$  & log(L/L$_{\odot})$ & e($_L$) & flag$_L$ & Mass & e$_{M [-1\sigma]}$ & e$({M [+1\sigma]})$
\\
\hline
  87 & 1.4248388 & 13.396266 & 5110 & 90 & -0.23 & -0.1 & 0.8 & 0 & 8.75 & 0.3 & 1.89 & 0.03 & 8.214 & 0.1339 & 1.96 & 0.03 & 0 & 2.99 & 0.5 & 0.01\\
  645 & 2.6784778 & -12.579897 & 4980 & 130 & -0.28 & -0.1 & 0.0 & 0 & 13.33 & 0.35 & 1.43 & 0.03 & 13.881 & 0.0803 & 1.40 & 0.02 & 0 & 1.9 & 0.4 & 0.6\\
  1013 & 3.6506853 & 20.206701 & 3600 & 150 & -2.24 & 0.0 & -1.1 & -1 & 8.86 & 0.22 & 2.98 & 0.02 & 9.1895 & 0.4342 & 3.01 & 0.06 & 0 & 1.52 & 0.71 & 1.46\\
  1483 & 4.6773424 & -43.23557 & 4420 & 110 & -0.58 & -0.1 & 0.0 & 0 & 8.89 & 0.49 & 1.70 & 0.05 & 7.3177 & 0.0392 & 1.88 & 0.02 & 0 & 1.18 & 0.31 & 0.67\\
  2261 & 6.571046 & -42.30598 & 4770 & 250 & -0.37 & -0.4 & 0.2 & 0 & 38.5 & 0.73 & 1.92 & 0.02 & -9999.0 & -9999.0 & 1.93 & 0.02 & 1 & 1.57 & 0.49 & 1.16\\
  2529 & 7.179697 & -50.53286 & 4650 & 110 & -0.43 & -0.1 & 0.3 & 0 & 7.98 & 0.43 & 1.76 & 0.05 & 8.4914 & 0.0322 & 1.72 & 0.02 & 0 & 1.41 & 0.46 & 1.08\\
  3712 & 10.126836 & 56.53733 & 4510 & 90 & -0.51 & -0.4 & 0.0 & 0 & 14.29 & 0.15 & 2.89 & 0.01 & 1.6949 & 1.2847 & 4.38 & 0.68 & 0 & 5.92 & 0.0 & 0.01\\
  3750 & 9.966457 & -44.79629 & 4610 & 110 & -0.45 & 0.0 & 2.5 & 0 & 10.62 & 0.43 & 1.63 & 0.04 & 10.2314 & 0.0489 & 1.67 & 0.02 & 0 & 1.49 & 0.5 & 1.0\\
  4042 & 11.012434 & 70.823425 & 5020 & 120 & -0.26 & -0.3 & 1.1 & 0 & 6.8 & 0.32 & 1.64 & 0.04 & 6.9952 & 0.0288 & 1.62 & 0.02 & 0 & 2.0 & 0.9 & 0.5\\
  4211 & 11.050414 & -38.421684 & 4600 & 100 & -0.46 & 0.0 & 0.0 & 0 & 9.63 & 0.4 & 1.76 & 0.04 & 10.1675 & 0.0691 & 1.72 & 0.02 & 0 & 1.54 & 0.5 & 0.96\\
  4229 & 9.914577 & -85.700874 & 4340 & 90 & -0.64 & 0.0 & 0.0 & -1 & 6.94 & 0.32 & 1.75 & 0.04 & 7.0051 & 0.0235 & 1.75 & 0.023 & 0 & 1.09 & 0.26 & 0.71\\
  4496 & 11.605771 & -57.927643 & 4510 & 90 & -0.51 & 0.0 & 0.0 & 0 & 6.71 & 0.57 & 1.56 & 0.08 & 4.912 & 0.0522 & 1.83 & 0.03 & 0 & 1.44 & 0.43 & 1.05\\
  4585 & 11.930084 & -18.061338 & 4290 & 110 & -0.7 & -0.2 & -0.5 & -1 & 9.18 & 0.41 & 1.97 & 0.04 & 8.8769 & 0.1143 & 2.03 & 0.03 & 0 & 1.02 & 0.21 & 0.64\\
  5316 & 13.811135 & 24.557062 & 3400 & 150 & -3.2 & -0.4 & -1.3 & -1 & 5.75 & 0.42 & 3.18 & 0.07 & 4.5201 & 0.1855 & 3.40 & 0.05 & 0 & 1.31 & 0.6 & 1.66\\
  5395 & 14.166271 & 59.181057 & 5030 & 260 & -0.26 & -0.4 & 0.0 & 0 & 16.32 & 0.23 & 1.73 & 0.02 & 17.2875 & 0.188 & 1.72 & 0.03 & 0 & 1.9 & 0.89 & 0.6\\
  5457 & 13.751303 & -69.527084 & 4680 & 90 & -0.42 & 0.0 & 0.0 & 0 & 14.7 & 0.27 & 1.55 & 0.02 & 14.6214 & 0.0817 & 1.58 & 0.02 & 0 & 1.59 & 0.65 & 0.9\\
  5722 & 14.682782 & -11.379975 & 5100 & 120 & -0.24 & -0.2 & 0.0 & -1 & 10.2 & 0.53 & 1.73 & 0.05 & 9.7623 & 0.112 & 1.79 & 0.028 & 0 & 2.49 & 0.56 & 0.5\\
  6559 & 16.532099 & -23.99245 & 4730 & 90 & -0.39 & -0.1 & 0.0 & 0 & 8.52 & 0.44 & 1.75 & 0.05 & 8.8661 & 0.0578 & 1.72 & 0.02 & 0 & 1.59 & 0.75 & 0.9\\
...&  ...& ... &... & ... &...  &... & ... &...  & ...& ...&...&...&...&...&...&...&...&...&...&...\\
\hline
\end{tabular}
}
\label{Observations_table}
  \end{center}
\end{table}
\end{landscape}

\begin{landscape}
\begin{table}
\caption{Informations on giants stars in three other samples used in this study. We use HD number to identify stars observed by \citet{Lucketal07} and \citet{Liu_etal14} while the Gaia identify is used for the GALAH sample.}
\label{othersample}
\centering 
\begin{tabular}{|r|r|r|r|r|r|r|r|r|}
\hline              
Name &  T$_{\rm{eff}}$ & $\pi_{\gaia}$ & e($\pi_{\gaia}$)  & log(L/L$_{\odot})$ & e($L$) & Mass & e(${M [-1\sigma]}$) & e$({M [+1\sigma]})$ \\
\hline
\multicolumn{9}{c}{\citet{Lucketal07}}\\
\hline
  448 & 4840 & 10.9923 & 0.1299 & 1.74 & 0.03 & 2.49 & 1.1 & 0.5\\
  3411 & 4657 & 8.7765 & 0.0385 & 1.71 & 0.02 & 2.49 & 1.28 & 0.5\\
  3546 & 5102 & 19.4949 & 0.2307 & 1.71 & 0.03 & 2.0 & 1.09 & 0.5\\
  4732 & 5021 & 18.2248 & 0.0565 & 1.13 & 0.02 & 1.9 & 0.3 & 0.1\\
  5722 & 4995 & 9.7623 & 0.112 & 1.80 & 0.03 & 2.49 & 0.84 & 0.5\\
  6037 & 4669 & 10.688 & 0.044 & 1.42 & 0.02 & 1.59 & 0.6 & 0.9\\
  6186 & 4955 & 17.2961 & 0.3842 & 1.87 & 0.04 & 2.49 & 0.95 & 0.5\\
  6559 & 4773 & 8.8661 & 0.0578 & 1.70 & 0.02 & 2.0 & 1.15 & 0.99\\
  7106 & 4748 & 19.2161 & 0.2278 & 1.71 & 0.03 & 1.9 & 0.9 & 1.09\\
  7578 & 4715 & 9.3814 & 0.0538 & 1.70 & 0.02 & 2.49 & 1.4 & 0.5\\
  8512 & 4771 & 28.6441 & 0.413 & 1.72 & 0.03 & 1.79 & 0.94 & 1.19\\
  ...&  ...& ... &... & ... &...  &... & ... &...  \\
  \hline
\hline
\multicolumn{9}{c}{\citet{Liu_etal14}}\\
\hline
  87 & 5072 & 8.214 & 0.1339 & 1.97 & 0.03 & 2.99 & 0.5 & 0.01\\
  360 & 4850 & 8.9376 & 0.0836 & 1.74 & 0.03 & 2.48 & 1.5 & 0.51\\
  448 & 4780 & 10.9923 & 0.1299 & 1.75 & 0.03 & 2.48 & 1.46 & 0.51\\
  587 & 4893 & 17.3901 & 0.1339 & 1.21 & 0.02 & 1.7 & 0.5 & 0.3\\
  645 & 4880 & 13.881 & 0.0803 & 1.41 & 0.02 & 2.0 & 0.5 & 0.5\\
  1239 & 5114 & 4.7303 & 0.0955 & 2.36 & 0.03 & 3.98 & 1.0 & 0.02\\
  1367 & 4982 & 8.875 & 0.0419 & 1.65 & 0.02 & 2.49 & 0.7 & 0.5\\
  1419 & 4888 & 8.5872 & 0.1468 & 1.73 & 0.03 & 2.49 & 0.6 & 0.5\\
  2114 & 5230 & 5.4912 & 0.0949 & 2.20 & 0.03& 2.99 & 0.01 & 1.0\\
  2952 & 4844 & 9.054 & 0.0535 & 1.75 & 0.02 & 2.49 & 1.39 & 0.5\\
  3421 & 5287 & 3.5023 & 0.1105 & 2.73 & 0.04 & 3.97 & 0.01 & 0.99\\
  ...&  ... & ... &...  &... & ... &...  & ...& ...\\

\hline
\multicolumn{9}{c}{GALAH \citet{GALAH2018}}\\
\hline
  4634991546064263040 & 4064 & 0.255 & 0.0193 & 2.55 & 0.08 & 1.24 & 0.44 & 1.24\\
  4644232799560251776 & 4052& 0.2225 & 0.0243 & 2.43 & 0.11 & 1.0 & 0.23 & 0.87\\
  4642633461114384640 & 4176 & 0.1932 & 0.0271 & 2.29 & 0.14 & 0.92 & 0.15 & 0.93\\
  4622338434971333248 & 4388 & 0.3384 & 0.0196 & 1.93 & 0.07 & 0.98 & 0.2 & 0.55\\
  4622323316686362752 & 4065 & 0.2681 & 0.0262 & 2.57 & 0.10 & 1.46 & 0.63 & 1.26\\
  2956458403404947712 & 4461 & 0.1282 & 0.0167 & 2.36 & 0.13 & 1.11 & 0.34 & 1.36\\
  4827471673666050816 & 4188 & 0.2602 & 0.0192 & 2.26 & 0.08 & 1.02 & 0.21 & 0.83\\
  4800298412016784896 & 4291 & 0.2295 & 0.0208 & 2.19 & 0.10 & 1.65 & 0.76 & 1.34\\
  2890175650594498048 & 4050 & 0.1305 & 0.0227 & 2.99 & 0.17 & 2.47 & 1.55 & 1.49\\
  2890162456454924544 & 4434 & 0.1656 & 0.0271 & 2.45 & 0.16 & 2.99 & 1.29 & 1.98\\
  2893967522602401792 & 4391 & 0.2155 & 0.0234 & 2.24 & 0.11 & 2.48 & 1.38 & 1.5\\
  ...&  ...& ... &... & ... &...  &... & ... &...  \\
  \hline
\end{tabular}
\end{table}
\end{landscape}

\begin{table*}
\caption{Information about LiRG from our sample that were identified as having a LTE Li abundance higher by +0.2~dex than prediction by standard models at the corresponding effective temperature. 
}
\centering
\begin{tabular}{|r|r|r|r|r|r|r|r|r|r|}
\hline
  \multicolumn{1}{|c|}{HD} &
  \multicolumn{1}{c|}{T$_{\rm{eff}}$} &
  \multicolumn{1}{c|}{e\_T$_{\rm{eff}}$} &
  \multicolumn{1}{c|}{log (L/L$_{\odot}$)} &
  \multicolumn{1}{c|}{e\_L} &
  \multicolumn{1}{c|}{[Fe/H]} &
  \multicolumn{1}{c|}{A(Li)} &
  \multicolumn{1}{c|}{Mass} &e$_{M [-1\sigma]}$ & e$_{M [+1\sigma]}$ \\
  \multicolumn{1}{|c|}{} &
  \multicolumn{1}{|c|}{K} &
  \multicolumn{1}{|c|}{K} & 
  \multicolumn{1}{|c|}{} &
  \multicolumn{1}{|c|}{} & 
  \multicolumn{1}{|c|}{dex} & 
  \multicolumn{1}{|c|}{dex} &
  \multicolumn{1}{|c|}{M$_{\odot}$} &
  \multicolumn{1}{|c|}{M$_{\odot}$} &
  \multicolumn{1}{|c|}{M$_{\odot}$} \\
\hline
 3750 & 4610 & 110 & 1.67 & 0.02 & 0.0 & 2.5 & 1.49 & 0.5 & 1.0\\
  4042 & 5020 & 120 & 1.62 & 0.026 & -0.3 & 1.1 & 2.0 & 0.9 & 0.5\\
  30197 & 4490 & 110 & 1.75 & 0.04 & 0.2 & 1.5 & 1.68 & 0.62 & 1.3\\
  40827 & 4620 & 110 & 1.83 & 0.02 & -0.1 & 1.9 & 1.57 & 0.53 & 0.93\\
  71129 & 4460 & 90 & 3.91 & 0.07 & -0.6 & 1.2 & 5.92 & 0.0 & 0.03\\
  77361 & 4600 & 90 & 1.87 & 0.02 & 0.0 & 4.6 & 1.78 & 0.59 & 1.21\\
  83506 & 4810 & 110 & 2.34 & 0.03 & -0.5 & 0.9 & 2.98 & 0.69 & 1.02\\
  85563 & 4530 & 90 & 1.76 & 0.04 & 0.0 & 1.5 & 1.39 & 0.4 & 1.1\\
  86634 & 4650 & 100 & 1.74 & 0.02 & 0.0 & 2.0 & 1.63 & 0.64 & 1.03\\
  90507 & 5030 & 110 & 1.48 & 0.02 & -0.3 & 1.3 & 1.9 & 0.4 & 0.6\\
  90633 & 4580 & 90 & 1.68 & 0.02 & 0.0 & 2.0 & 1.42 & 0.43 & 1.07\\
  93859 & 4590 & 110 & 1.92 & 0.03 & -0.3 & 0.9 & 1.39 & 0.4 & 1.09\\
  106574 & 4490 & 90 & 2.19& 0.03 & -0.4 & 1.1 & 1.53 & 0.55 & 0.96\\
  113049 & 4810 & 110 & 2.17 & 0.031 & -0.3 & 1.1 & 2.98 & 1.18 & 1.02\\
  115299 & 4630 & 110 & 1.71 & 0.02 & -0.2 & 3.6 & 1.29 & 0.34 & 0.6\\
  133086 & 4890 & 110 & 1.73 & 0.02 & -0.2 & 1.9 & 2.0 & 1.05 & 0.5\\
  138289 & 4460 & 100 & 1.72 & 0.02 & 0.2 & 3.0 & 1.39 & 0.41 & 1.1\\
  183912 & 4680 & 80 & 2.96 & 0.07 & -0.1 & 1.7 & 4.99 & 1.0 & 0.97\\
  188114 & 4810 & 250 & 1.98 & 0.04 & -0.4 & 0.9 & 1.87 & 0.75 & 1.11\\
  199437 & 4560 & 110 & 2.18 & 0.03 & -0.3 & 0.9 & 1.85 & 0.66 & 1.13\\
  206078 & 4940 & 80 & 1.58 & 0.02 & -0.6 & 1.1 & 1.49 & 0.61 & 0.4\\
\hline\end{tabular}
\label{Table:Li-rich_oursample}
\end{table*}

\begin{table*}
\caption{Information about LiRG that were identified as having NLTE Li abundance higher by +0.2~dex than prediction by standard models at corresponding effective temperature in the sample of \citet[][]{Lucketal07} and \citet[][]{Liu_etal14}. 
}
\label{Table:Li-rich_othersamples}
\centering 
\begin{tabular}{|r|r|r|r|r|r|r|r|r|}
\hline 
  \multicolumn{1}{|c|}{HD} &
  \multicolumn{1}{c|}{T$_{\rm{eff}}$} &
  \multicolumn{1}{c|}{log (L/L$_{\odot}$)} &
  \multicolumn{1}{c|}{e\_L} &
  \multicolumn{1}{c|}{[Fe/H]} &
  \multicolumn{1}{c|}{A(Li)} &
  \multicolumn{1}{c|}{Mass} &e$_{M [-1\sigma]}$ & e$_{M [+1\sigma]}$\\
  \multicolumn{1}{|c|}{} &
  \multicolumn{1}{|c|}{K} &
  \multicolumn{1}{|c|}{} &
  \multicolumn{1}{|c|}{} & 
  \multicolumn{1}{|c|}{dex} & 
  \multicolumn{1}{|c|}{dex} &
  \multicolumn{1}{|c|}{M$_{\odot}$} &
  \multicolumn{1}{|c|}{M$_{\odot}$} &
  \multicolumn{1}{|c|}{M$_{\odot}$} \\
  \hline
\multicolumn{9}{c}{\citet{Lucketal07}}\\
\hline
  186815 & 5116 & 1.32 & 0.02 & -0.22 & 1.29 & 1.9 & 0.2 & 0.1\\
  214995 & 4737 & 1.55 & 0.02 & 0.0 & 3.01 & 1.59 & 0.75 & 0.9\\
\hline
\multicolumn{9}{c}{\citet{Liu_etal14}}\\
\hline
  5395 & 4774 & 1.75 & 0.03 & -0.45 & 1.3 & 1.39 & 0.54 & 0.5\\
  35410 & 4809 & 1.59 & 0.03 & -0.33 & 1.26 & 1.39 & 0.55 & 0.6\\
  102845 & 4975 & 2.01 & 0.03 & -0.07 & 2.11 & 2.99 & 0.5 & 0.01\\
  138905 & 4822 & 1.85 & 0.03 & -0.3 & 1.44 & 1.79 & 0.79 & 1.19\\
  160781 & 4593 & 2.6 & 0.04 & -0.02 & 1.46 & 3.99 & 1.01 & 1.0\\
  196857 & 4878 & 1.70 & 0.02 & -0.27 & 1.22 & 1.69 & 0.83 & 0.8\\
\hline
\end{tabular}
\end{table*}

\begin{acknowledgements}
      Part of this work was supported by the French "Programme National de Physique Stellaire" (PNPS) and "Programme National Cosmologie et Galaxies" (PNCG) of CNRS/INSU, and by the Swiss National Science  Foundation (SNF).  RS acknowledges support by the National Science Center of Poland through grant 2012/07/B/ST9/04428 and support from the Polish Ministry of Science and Higher Education. 
  PN thanks Dr. J.Pritchard for his help with the FEROS pipeline. 
  We thank R.Gauderon for having observed with FEROS@ESO in March 2004.
  JKH and observations taken at McDonald Observatory were supported by the McDonald REU program under NSF AST-0649128. We thank the anonymous referee for important comments that helped improve the paper.
\end{acknowledgements}

\bibliographystyle{aa}
\bibliography{Charbonnel_etal_LiGiants_biblio.bib}

\end{document}